\documentclass[sn-basic]{sn-jnl}

\usepackage{graphicx}%
\usepackage{multirow}%
\usepackage{amsmath,amssymb,amsfonts}%
\usepackage{amsthm}%
\usepackage{mathrsfs}%
\usepackage{appendix}%
\usepackage{xcolor}%
\usepackage{textcomp}%
\usepackage{manyfoot}%
\usepackage{booktabs}%
\usepackage{algorithm}%
\usepackage{algorithmicx}%
\usepackage{algpseudocode}%
\usepackage{listings}%
\usepackage{ulem}
\usepackage[latin1]{inputenc}

\theoremstyle{thmstyleone}%
\theoremstyle{thmstyletwo}%

\theoremstyle{thmstylethree}%

\newcommand{\nl}{\newline}
\newcommand{\comment}[1]{}

\begin{document}


\title[Article Title]{The birth of StatPhys: The 1949 Florence conference at the juncture of national and international physics reconstruction after World War II}

\author*[1,2]{\fnm{Roberto} \sur{Lalli}}\email{roberto.lalli@polito.it}

\author[3,4]{\fnm{Paolo} \sur{Politi}}\email{Paolo.Politi@cnr.it}
\equalcont{This author contributed equally to this work.}

\affil*[1]{\orgdiv{Department of Mechanical and Aerospace Engineering}, \orgname{Politecnico di Torino}, \orgaddress{\street{Corso Duca degli Abruzzi 24}, \city{Torino}, \postcode{10129}, \country{Italy}}}

\affil[2]{\orgdiv{Department I}, \orgname{Max Planck Institute for the History of Science}, \orgaddress{\street{Boltzmannst. 22}, \city{Berlin}, \postcode{14195}, \country{Germany}}}

\affil[3]{\orgdiv{Istituto dei Sistemi Complessi}, \orgname{Consiglio Nazionale delle Ricerche}, 
\orgaddress{\street{Via Madonna del Piano 10}, \city{Sesto Fiorentino}, \postcode{50019}, \country{Italy}}}

\affil[4]{\orgdiv{Istituto Nazionale di Fisica Nucleare}, \orgname{Sezione di Firenze},
\orgaddress{\street{Via G. Sansone 1}, \city{Sesto Fiorentino}, \postcode{50019}, \country{Italy}}}


\abstract{In spring 1949 about 70 physicists from eight countries met in Florence to discuss recent trends in statistical mechanics. This scientific gathering, co-organized by the Commission on Thermodynamics and Statistical Mechanics of the International Union of Pure and Applied Physics (IUPAP) and the Italian Physical Society (SIF), initiated a tradition of IUPAP-sponsored international conferences on statistical mechanics that lasts to this day. In 1977, when this conference series took the name of StatPhys, the foundational role of the Florence conference was recognized by retrospectively naming it StatPhys1. This paper examines the dual scientific and social significance of the conference, situating it in the broader contexts of the post-World War II reconstruction in Italian physics and of the revitalization of the international science organization. Through an analysis of IUPAP archives and Italian records, we illustrate how the event's success hinged on the aligned objectives of its organizers. Internationally, it was instrumental in defining the scientific and organizational foundations for the activities of IUPAP commissions during a critical phase of IUPAP's history, when the Union was resurging on the international scene after the inactivity of the interwar period. Nationally, the conference served as a cornerstone in SIF's strategy to re-establish Italian physics' international stature and to aid the domestic revitalization of physics through the internationalization of its activities, notably of its flagship journal, \textit{Il Nuovo Cimento}. This analysis not only sheds light on the conference's impact but also informs recent discussions in the history of science about the multiple roles of international scientific conferences.}




\maketitle

\section{Introduction}\label{sec1}

\begin{figure}[h]%
\centering
\includegraphics[width=0.9\textwidth]{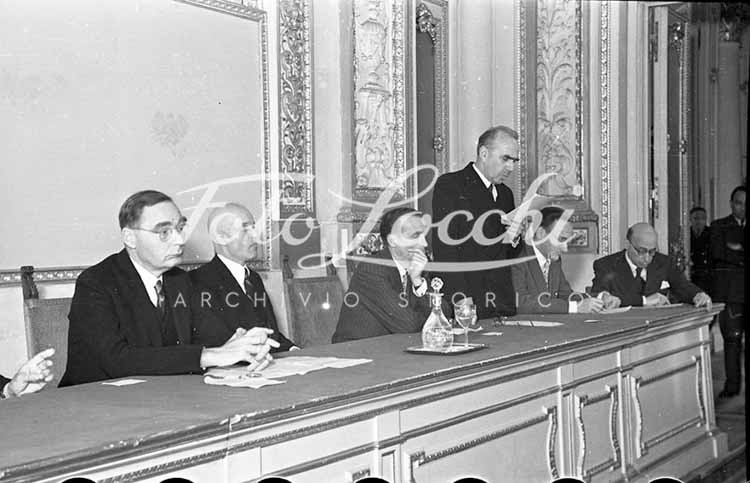}
\caption{Opening session at Villa Favard on May 17, 1949. From left to right: Hendrik Kramers (IUPAP president and co-chairman of the conference), Bruno Borghi (Rector of the University of Florence), Mario Fabiani (Mayor of Florence), Giovanni Polvani (standing, president of SIF and co-chairman of the conference), Giuliano Toraldo di Francia.
$\copyright$ Foto Locchi.}
\label{Locchi_podio}
\end{figure}

\begin{figure}[h]%
\centering
\includegraphics[width=0.9\textwidth]{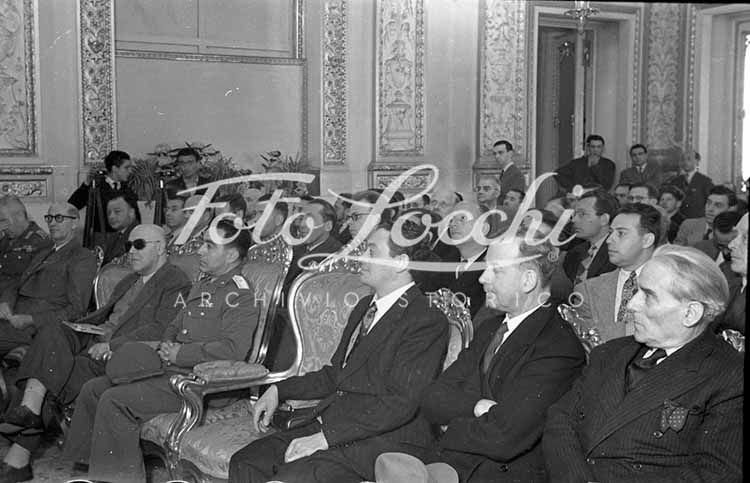}
\caption{Audience at the opening session at Villa Favard on May 17. 
We can recognize Pierre Fleury (General Secretary of IUPAP,  first row, second from the left), 
Wolfgang Pauli (second row, first on the left), 
L\'eon van Hove (second row, seventh on the left), 
Max Born (second row, first on the right),
Giacomo Morpurgo (second row, second on the right), 
E.G.D. Cohen (fourth row, first on the right).
$\copyright$ Foto Locchi.}
\label{Locchi_platea}
\end{figure}

On May 17\textsuperscript{th} 1949, about 70 physicists\footnote{Almost all males, how often occurred 
in that historical period for these sorts of international scientific gatherings.} 
from the United States and Europe, almost uniquely Western, 
met at Villa Favard  in Florence for the opening session of a conference devoted to discuss frontier research topics from the statistical mechanics of interacting systems (see Figs.~\ref{Locchi_podio} and \ref{Locchi_platea}).\footnote{Giovanni Polvani to Mons. Arnoux, March 24, 1949,  Box 3, Folder 2, Presidenza Polvani, Archives of the Italian Physical Society (hereafter ASIF).}
This 4-day conference initiated a regular series of increasingly larger international meetings dedicated to statistical physics that lasts to this day: the StatPhys international conferences sponsored by the International Union of Pure and Applied Physics (IUPAP). In the 1970s, when this tradition was well established and StatPhys emerged as the official name of these regular conferences, the role of the Florence conference was ratified by considering it StatPhys1.\footnote{In the documents of the IUPAP archives the first mention of `StatPhys' occurs in relation to the conference held in Haifa in 1977, called StatPhys13; see, e.g., \textit{IUPAP General Report 1979}, p. 54, Series B2aa, Vol. 2,  IUPAP Archives, Gothenburg Secretariat (hereafter IuG), Center for the History of Science, Royal Swedish Academy of Sciences; see also the proceedings of the conference \citep{cabibStatisticalPhysicsStatphys1978}. However, we do not exclude that the `StatPhys' label was informally used before the 1977 conference.}

As the first of an important disciplinary tradition, it played a foundational role in the international development of a major sub-field of physics, and thus warrants a deeper historical examination than it has received to date. Even more so when one recognizes that the conference was the result of two parallel re-construction processes in the post-World War II period: the re-establishment of the international organization of physics and the attempts to (re-)build Italian physics. It was a collaborative effort between the newly revitalized Italian Physical Society (SIF, from its Italian name \textit{Società Italiana di Fisica}) and the IUPAP Commission on Thermodynamics and Statistical Mechanics, which was notable for being the first specialized commission ever established by IUPAP, just two years prior. The conference's pioneering role underscores its significance not only as a platform that fostered the development of an international sub-disciplinary framework within the broader scientific discipline of physics but also as a means for the Italian scientific community to reassert its prominence in the international arena after the war's devastations. 

International scientific conferences have garnered increasing interest from historians of science  as key objects for exploring the cultural, social, and political dimensions of the scientific enterprise. The recent European project  \textit{The Scientific Conference: A Social, Cultural and Political History} has shed light on the vast diversity of these gatherings, introducing a taxonomy that categorizes them into four distinct types: disciplinary conferences, which cultivate communities within scientific fields and establish common standards; scientific associations' conferences primarily aimed at community building; technical conferences, which facilitate interaction between scientists and technicians in inter-governmental contexts to address technical or political challenges; and small `elite' conferences sponsored by wealthy philanthropists, aimed at reinforcing politically-charged research agendas \citep{biggArtGatheringHistories2023}.\footnote{See the website's project at https://heranet.info/projects/public-spaces-culture-and-integration-in-europe/the-scientific-conference-a-social-cultural-and-political-history/.} According to the project's team, despite their varied formats and objectives, these conferences share certain characteristics that render them indispensable to the scientific community. Drawing upon Émile Durkheim's analysis of religious gatherings, historian Geert Somsen argues that scientific conferences enact rituals and routines that have purely cultural and social goals. They enable participants to celebrate their collective identity and engage with a shared sense of purpose and solidarity~\citep{somsenGoddessThatWe2023}. 
Another important aspect highlighted by the project is that scientific conferences may have geopolitical goals too, as they were, and still are, recognized by larger institutions as valuable tools for fostering international relations~\citep{reinischTechnicalConferencesTechnique2023}. These insights provide a comprehensive framework for examining international scientific gatherings, revealing their multifaceted roles, norms, and practices. This approach offers a lens through which to view specific historical instances, such as the conference under discussion, but it also allows to articulate and complement that framework looking at a different historical case. In fact, the Florence conference presents a special case of the ``disciplinary conference" type of the above mentioned taxonomy, one mostly defined by its focus on a specific sub-discipline within a larger disciplinary field in relation to an institutional framework devoted to the entire discipline.

The 1949 conference was co-organized by IUPAP, the major international organization devoted to physics. Recently, there has been a growing interest in the study of scientific organizations within the framework of science diplomacy history.\footnote{For historiographical analyses of current historiographical trends on science diplomacy, see~\citep[e.g.,][]{turchettiIntroductionJustNeedham2020a,adamsonGlobalPerspectivesScience2021}.} The crucial public and military significance of physics immediately following World War II has highlighted IUPAP as a subject of particular interest for analysis through the lens of science diplomacy. Recent historical research has revealed the intricate interactions between scientific endeavors and political strategies that influenced IUPAP's operations during the early Cold War~\citep{lalliGlobalizingPhysicsOne2024}. These analyses are essential for understanding the broader international context surrounding the Florence conference. Among such studies, Japanese historian of science Kenji Ito~(\citeyear{itoRepairingScientificNetwork2024}) has illustrated the essential role international conferences played in the maintenance of knowledge infrastructure and in allowing local scientific communities to rebuild ties after the end of World War II. This was the case of the International Conference of Theoretical Physics held in 1953 Japan, which was pivotal for the local physics community to re-appropriate a central space in the international organization of science that was being rebuilt.

Basing our investigation on documents found in the recently digitized IUPAP archival collections and other archival collections in Italy, in this paper we argue that the Florence conference played a similar and even more pioneering role in an earlier phase of the Cold War for the Italian physics community as well for the emerging community of physicists interested in statistical mechanics. In relation to the recent studies on international scientific conference summarized above, we ask: what was the main scientific, cultural, social and political purposes for the organization of the Florence meeting? What kind of meeting was it? What role did it play in the general strategies and activities of the two co-organizing institutions at the international and national levels? How far were the goals achieved? What was its influence in later developments of statistical physics and the related scientific community? 

To discuss the multiple elements embedded in the organization of the conference, we organize the paper as follows. We start by sketching the prehistory of the event, first by summarizing the development of IUPAP in the interwar period in Section~\ref{sec2}, and, in Section~\ref{sec3}, by delving into the political issues characterizing Italian changing relations with IUPAP in that period. In Section~\ref{sec4}, we contextualize the foundation of the Commission on Thermodynamics and Statistical Mechanics in the post-World War II refoundation of IUPAP and of the international organization of science. In Section~\ref{sec5}, we discuss the Florence conference as a key moment in the international \mbox{(re-)construction} of Italian physics after World War II. In Section~\ref{sec6}, we analyze the scientific content of the conference and its role in the process of discipline-formation for an international community of statistical physicists. In Section~\ref{sec7}, we report some major transformations in the following developments of the Commission and its conferences up to the establishment of the name StatPhys13 for the 1977 conference held in Haifa. In the conclusion (Section~\ref{sec8}), we review the impact and role of the conference for both the international community of statistical physicists and for the local community of Italian physicists, discussing the implications of our case study for the cultural, social, political and scientific history of scientific conferences. This paper contributes to filling a historiographical gap in the 20th-century history of physics by exploring a significant episode in the development of statistical mechanics and statistical physics post-World War II, an area that has not been comprehensively investigated yet.

\section{IUPAP in the interwar period}\label{sec2}

To accurately contextualize the 1949 Florence conference, it is essential to consider both the general evolution of IUPAP during the interwar period and the specific dynamics of Italian participation, as these aspects were significantly influenced by the shifting political landscapes of the time. IUPAP was funded in 1922 within the International Research Council (IRC), the main framework for the re-organization of international science after World War I.\footnote{For the history of IUPAP in the interwar period see~\citep{fauqueIUPAPInterwarWorld2024,navarroHappyThirtiesMillikan2024}.} Like all scientific unions of the IRC family, the structure and membership of IUPAP was shaped by the punitive attitude of the victorious nations against the Central Powers. At the time of the IRC founding in 1919, the exclusion of countries belonging to the Central Powers was decreed by statute. A period of twelve years was fixed before this rule could be modified. Only when the original convention expired, in December 1931, would it be possible to amend the statutes and admit the countries initially excluded. These rules applied to all unions that operated within the IRC, as only institutions representing countries that were already members of the IRC could be admitted as members of the unions. This rule was based on the notion of national membership according to which the scientific members of both the IRC and the unions were official national representatives.\footnote{The institutional nature of national representatives varied greatly. For instance, a 1928 report reads: ``out of the thirty-five countries which have joined the International Research Council, fourteen are represented by their scientific academies, six by national research councils composed of representatives of the national academies, one by a scientific society, and seven others by a scientific department connected with its government. In seven cases only out of the thirty-five is the government the adhering body"~\citep[p. 390]{InternationalResearchCouncil1928a}.} 

At the initial General Assembly of the IRC in 1919, a preliminary statute for a Union of Physics was drafted, but it wasn't until 1922 that a temporary ten-member Executive Committee was formed to officially establish the Union.\footnote{``Union internationale de physique pure et appliquée. Procés-Verbal Assemblée Générale Constitutive, Paris, Décembre 1923,'' Series B2aa, Vol. 1,  IuG.} Relative to other scientific unions, such as the International Astronomical Union (IAU), IUPAP was marked by periods of inactivity and diminished influence during the interwar years. The limitations of an organization devoted to international cooperation based on the boycott of German scientists became evident especially in the field of physics, where major developments in theoretical physics had German-speaking scientific communities as their nerve center. Furthermore, the evolving diplomatic landscape, epitomized by the Locarno Pact of 1925 between France and Germany and Germany's subsequent admission into the League of Nations in 1926, highlighted the outdated and scientifically detrimental nature of excluding German scientists from the IRC and its unions, against the backdrop of the political détente of that era.

Although modifications to the IRC statutes in 1926 nominally allowed for the membership of League of Nations countries, potentially paving the way for Germany's inclusion in the IRC and its unions, the intransigent positions of French and Belgian scientists, on the one side, and of German academies, on the other, prevented this integration~\citep{schroeder-gudehusScientifiques1978a}. This deadlock had significant repercussions for IUPAP, which lagged in fostering full international cooperation with German physicists, in contrast to the broader trend of re-engagement seen in key meetings like the Como and Solvay conferences of 1927. IUPAP remained characterized by an extremely partial international cooperation despite the fact that William Bragg (president of IUPAP from 1922 to 1931), the French physicist Henri Abraham (IUPAP secretary general from 1922 until his death at Auschwitz in 1944), and other members of IUPAP's Executive Committee had argued in favor of an early entry of German physicists, even proposing to suspend assemblies until the full participation of German physicists was feasible.\footnote{``Union internationale de physique pure et appliquée. Deuxième assemblée générale. Bruxelles, 7 juillet 1925,'' Series B2aa, Vol. 1, IuG.} As a testament to this stance, IUPAP did not convene any conferences between its second General Assembly in 1925 and 1931, coinciding with the year when the IRC and its unions finally amended their statutes to facilitate broader membership.

When the IRC transformed into the International Council of Scientific Unions (ICSU) in 1931, there was a widespread anticipation that German scientific institutions would swiftly become members of ICSU or its individual unions. However, this integration did not materialize as expected. A significant portion of the German scientific community remained reluctant to join an international body that was originally established with the intention of excluding them. Compounding this issue, disagreements between German academies about which one legitimately represented the German state, along with the evolving economic and political landscape---marked by the aftermath of the 1929 stock market crash and the Nazi Party's ascendancy in 1933---further complicated Germany's potential membership in ICSU.

IUPAP resumed its activities in 1931 with the hope that German physicists would soon become part of the organization. At the General Assembly that year, U.S. physicist and Nobel Laureate, Robert Millikan, became president of IUPAP with this ambition in mind and grand plans for the extension of the IUPAP activities, including the establishment of the first general scientific commissions, the most relevant of which was the Commission on Symbols, Units and Nomenclature (SUN) aimed to establish international standards. When the anticipated integration of German physicists failed to materialize, these initiatives were hindered from the beginning. A clear sign of these difficulties was that IUPAP didn't even have a president from 1934 to 1937. After learning that he had been elected in absentia at the IUPAP 1934 General Assembly as Millikan's successor, Niels Bohr declined the position, stating that he could not be president of an organization that did not embody the principle of being ``truly international."\footnote{N. Bohr to H. Abraham, December 12, 1934, Series E1, Vol. 5, Folder 38, IUPAP Archives, Quebec Secretariat (hereafter IuQ), Center for the History of Science, Royal Swedish Academy of Sciences.} In 1937, Enrico Fermi also declined the position without providing any reasons.\footnote{Fermi to Abraham, October 4, 1937, Series E1, Vol. 5, Folder 38, IuQ.}\footnote{It is useful to note that the 1937 offer to Enrico Fermi to become president of IUPAP demonstrates his high international standing at the time, confirmed by the awarding of the Nobel Prize one year later. It also shows that Fermi, like Bohr and Siegbahn, was considered close enough to the German physics community to be able to involve them in IUPAP activities.} Only later did Swedish physicist and Nobel Laureate Manne Siegbahn become the new president. At the time, Siegbahn was establishing the first research institute for nuclear physics in Sweden, a country with strong cultural and scientific ties with Germany. He became president still hopeful of engaging German physicists, but the prevailing political conditions thwarted any significant revival of IUPAP's endeavors before the onset of World War II brought its activities to a complete standstill.

\section{The Italian participation in IUPAP in the interwar period}\label{sec3}

Let's now examine how Italian scientists and politicians interacted with IUPAP between the two World Wars contextualizing it within the overarching state of affairs previously outlined. In the foundational phase of the IRC, given Italy's position as one of the Allies against the Central Powers, Italian scientists played a pivotal role in the negotiations that defined the organization's structure. A key figure in these discussions was physicist and mathematician Vito Volterra, who was appointed as one of the five members of the IRC Executive Committee in 1919, serving in the capacity of vice-president.

Volterra, Senator of the Italian Kingdom since 1905, was one of the most active figures in organizing Italian science and promoting it abroad. His engagement with the IRC was deeply intertwined with his and other Italian scientists' ambitions to utilize the international institutional landscape as a means to reform the organization of scientific and technological research in Italy. In the lead-up to the establishment of the IRC, it was envisioned that each member country would set up  a central state body to organize research on the model of the National Research Council, created in the United States in 1916. 
Volterra championed this project in Italy, initiating a process that would lead to the creation of the National Research Council (CNR, from its Italian name \textit{Consiglio Nazionale delle Ricerche}), established in November 1923 within the \textit{Accademia dei Lincei}, with Volterra being the chairman of both the CNR and the \textit{Accademia dei Lincei}~\citep{tomassiniOrigini2001}. Prior to the CNR's inception, the \textit{Accademia dei Lincei} served as Italy's primary representative in international scientific fora. As president of the \textit{Accademia dei Lincei},  Volterra played a pivotal role in orchestrating Italy's involvement in IUPAP, contributing to the drafting of the preliminary statutes in 1919.\footnote{Volterra's personal commitment in the organization of the Italian physics community is attested by his status as one of the founding members of SIF in 1897, where he served as president from 1907 to 1909.} He was also instrumental in nominating physicist Orso Mario Corbino, a fellow member of the \textit{Accademia dei Lincei} and a Senator of the Kingdom, to the temporary ten-member IUPAP Executive Committee in July 1922.\footnote{Corbino was also a leading figure in the revival of Italian physics in the late 1920s. One of his major accomplishments was recognizing that Enrico Fermi could play a key role in spreading modern physics in Italy and educating new generations. With support from influential mathematicians like Volterra, Corbino facilitated the establishment of Italy's first chair of theoretical physics, to which Fermi was appointed in 1926, see~\citep{focacciaOrsoMarioCorbino2022}.}

The interval between the establishment of IUPAP's Executive Committee and its inaugural General Assembly in December 1923 coincided with dramatic transformations in the Italian political landscape. The march on Rome by the National Fascist Party's militants in October 1922 culminated in a regime change, with Benito Mussolini being appointed to form a new government. Although avowedly anti-fascist, Volterra initially collaborated with the new regime on matters concerning the country's scientific and technological development and the creation of the CNR, which was finalized during the Mussolini government. The CNR's bylaws mandated its affiliation with the IRC and called for the establishment of specialized commissions to integrate into the various unions under the IRC umbrella~\citep{guerraggioVitoVolterra2008}. At the time of IUPAP's formation, however, the CNR and these commissions had not been established yet. In fact, at the first IUPAP General Assembly Volterra was the only Italian representative. It wasn't until December 1925, following the second IUPAP General Assembly and after the institutional limitations and the impact of national political changes on Italian engagement had become evident, that a physics committee within the CNR was constituted.
 
The consolidation of the Fascist government into a dictatorship in 1925 precipitated the reformation of the CNR in 1927, leading to its detachment from the \textit{Accademia dei Lincei} and its reconstitution as a governmental entity. In this new structure, the directorate was nominated by the government and reported directly to the Head of the Italian Government, namely, Benito Mussolini. This reform also resulted in the dismissal of Volterra from his roles within the CNR, with Guglielmo Marconi, who was more aligned with the Fascist regime, being installed as the CNR's president.\footnote{Later, after Volterra refused to take the oath of loyalty to Mussolini's government in 1931, he was forced to leave the University of Rome and resign from all Italian scientific academies, where he had long been one of the most distinguished members.} Despite this change, during the 1928 IRC assembly, Volterra's position as vice-president of the IRC was reaffirmed.  This decision faced opposition from the newly appointed Italian representatives of the CNR, who contested Volterra's election on the grounds that he no longer served as an official Italian delegate, thus deeming his election illegal.\footnote{Magrini to Arthur Schuster, September 5, 1928,  Box 20, Folder 366, Fondo Consiglio Nazionale delle Ricerche, serie Presidenza Marconi (hereafter CNRMarconi), Archivio Centrale dello Stato, Rome (hereafter ACS); see also \citep{paoloniOrganizzazioneRisorseDi2001}.} In response to this objection, the IRC Executive Committee staunchly defended the decision, asserting that Volterra's election was fully compliant with IRC regulations.\footnote{Arthur Schuster to Magrini, August 2, 1928, Box 20, Folder, 366, CNRMarconi.}

This scenario prompted Italy to adopt stringent positions against the IRC's framework. In 1928, the IRC initiated discussions to revise its statutes, aiming for an organizational overhaul in 1931 that would potentially enable Germany and other previously excluded nations to join. The CNR advocated forcefully for the unions' complete independence from the IRC, a stance largely driven by the desire to dissociate from anti-fascist sentiments perceived to be prevalent within the IRC's leadership, as stressed by the vice-president of the CNR, jurist and diplomat Amedeo Giannini.\footnote{A. Giannini to the secretary of the IRC, Sir H. Lyons, August 14, 1929, Box 20, Folder 367, CNRMarconi; ``Verbale della seduta del direttorio del Consiglio Nazionale delle Ricerche, 31 Ottobre 1930,'' Archives of CNR, Roma (hereafter ACNR).}  By 1939, with the discussions on revising the IRC still underway, the CNR directorate started to evaluate the Italian involvement in the individual unions. It solicited reports from the Italian commissions assessing the performance of various unions ``because it is known that some unions work well, some mediocrely, and others do not work at all."\footnote{``Verbale della seduta del direttorio del Consiglio Nazionale delle Ricerche," October 31, 1929, ACNR, all translations from Italian documents are by the authors.}  Among the unions considered unnecessary by the directorate of the CNR there certainly was IUPAP, to which the CNR had not paid membership dues since 1927.

The deliberations on Italian participation in the IRC became matter of Italian foreign policy, with the aim of positioning Italy at the vanguard of an international campaign to boycott the IRC. It was concluded that ``the [International Research Council] as it is organized serves no purpose," leading the directorate to recommend to Mussolini that Italy withdraw from the IRC for ``essentially political" reasons.\footnote{Promemoria, undated, Box 20, Folder 266, CNRMarconi; ``Verbale della seduta del direttorio del Consiglio Nazionale delle Ricerche," January 22, 1931; and ``Verbale della seduta del direttorio del Consiglio Nazionale delle Ricerche," February 13, 1931, ACNR.} Beyond these political considerations, the directorate cited the proliferation of international organizations in which Italy held membership as overly burdensome. They suggested a more selective approach whereby Italy would engage with only one scientific organization per discipline, chosen with careful consideration of the organization's quality and utility. It is not difficult to see that the CNR directorate did not regard IUPAP as meeting these criteria, viewing participation in such unuseful unions as not only unnecessary but also ``harmful."\footnote{Promemoria, undated, Box 20, Folder 266, CNRMarconi.}
 
This strategic decision was bolstered by a political campaign aimed at informing, through Italian ambassadors, all the countries that had previously been excluded or were still undecided about affiliating with the successor to the IRC, with a particular focus on Germany.\footnote{Marconi to the Ministry of Foreign Affairs, February 5, 1931, Box 20, Folder 366, CNRMarconi; beyond Germany, the States informed of the Italian position about the IRC membership were: Argentina, Austria, Bulgaria, Brazil, Hungary, Spain, Turkey. See, Sottosegretario di Stato to G. Marconi, October 14, 1931, Box 20, Folder 366, CNRMarconi.} The goal was to publicly declare Italy's stance on boycotting the IRC, thereby paving the way for other unaffiliated countries to sustain their dissociation from the IRC beyond 1931. This strategy aimed at potentially fostering a new framework for the organization of international science, with Italy assuming a leading role.
Consequently, the transition from the IRC to ICSU in 1931 unfolded without Italy's participation. And the same happened for IUPAP. At the 1931 General Assembly of the Union Italy was no longer among the member countries, while at the same time continuing its participation in the activities of other unions.\footnote{``Union internationale de physique pure et appliquée. Troisième assemblée générale. Bruxelles, juillet 1931,'' Serie B2aa, Vol. 1, IuG; for the continued participation in other unions, see sub-folder Mathematical Union, Box 20, Folder 367, Sub-folder 8, CNRMarconi.}

Many scientists involved in the transition from the IRC to ICSU viewed, instead, the new organizational structure as an opportunity to re-establish international cooperation on new terms, potentially paving the way for German inclusion. When Robert Millikan assumed the presidency of IUPAP in 1931, he was particularly concerned about Italy's absence from the organization. He personally engaged in discussion with Marconi to encourage Italian re-engagement with IUPAP. Following their conversation at the nuclear physics congress in Rome in October 1931, Millikan did not wait to return to the United States before reaching out to Marconi and seeking assurances of Italy's full participation in IUPAP's activities.\footnote{Millikan to Marconi, November 25, 1931, Roll 12, Robert A. Millikan Papers [microfilm], Caltech Archives (hereafter Millikan Papers); see, also~\citep{navarroHappyThirtiesMillikan2024}.}

These developments, combined with the impression that Germany was on the verge of joining ICSU and its unions, spurred a strategic reevaluation within the CNR directorate. In discussions with Mussolini, a consensus emerged that Italy should actively seek membership in ICSU and promptly inform all countries previously notified of Italy's disengagement.\footnote{G. Marconi to the Minister of Foreign Affairs, May 12, 1931, Box 20, Folder 366, CNRMarconi.}  
After this decision, Italy rejoined IUPAP, even paying the back dues. This pivot resulted in Italy's reintegration into IUPAP, inclusive of settling outstanding dues.\footnote{``Union internationale de physique pure et appliquée. Quatrième assemblée générale. Londres, 5 octobre 1934,'' Serie B2aa, Vol. 1, IuG.} Consequently, Italy's membership appeared uninterrupted, although our analysis reveals that, for a period, the CNR's strategy was to maintain affiliations solely with selected unions deemed most beneficial, which did not include IUPAP. Italy's engagement with IUPAP remained consistent up until the outbreak of World War II. However, IUPAP's failure to formally incorporate German scientists and its subsequent inactivity led to a perception of Italy's involvement as being in a state of perpetual stagnation, mirroring that of IUPAP itself. This view is further highlighted by Enrico Fermi's choice to decline the presidency of IUPAP following Niels Bohr's refusal, signaling the organization's unattractiveness to Italian physicists as a platform for international scientific collaboration.

\section{The foundation of the IUPAP Commission on Thermodynamics and Statistical Mechanics in the post-World War II period}\label{sec4}

After World War II, IUPAP was reestablished on entirely new foundations.\footnote{For a detailed discussion of IUPAP's refoundation after World War II, see \citep{lalliDiplomacyPhysicsBack2024}.} During the fifth IUPAP General Assembly in January 1947, nearly forty physicists convened in Paris to deliberate on the organization's future direction. Reflecting on IUPAP's shortcomings in the interwar period, the acting IUPAP secretary general, UK-based German physicist Paul P. Ewald, underlined the necessity for a transformative approach in the organization's activities. He advocated for principles of inclusivity and openness to all nations, including those that had been defeated during World War II ``as soon as political conditions would permit."\footnote{``IUPAP, Report 5th General Assembly, September 1947," p. 17, Series B2aa, Vol. 1, IuG, translation by the authors.} 

In conjunction with these individual physicists' bottom-up efforts to reconstitute IUPAP around the ethos of open international cooperation, a new system of international scientific institutions emerged at the end of World War II. This system was intricately connected with the establishment of the United Nations as the main organization in the architecture of the new global order as well as with the legal differentiation between intergovernmental and non-governmental scientific institutions delineated in the 1946 UN Charter. In December 1946, ICSU had signed a formal agreement with UNESCO, the UN agency devoted to education, science, and culture. This agreement not only ensured increased funding opportunities for ICSU and its affiliated unions but also conferred upon them the status of being the principal non-governmental organizations recognized by the United Nations for all affairs related to the natural sciences~\citep{petitjean_sixty_2006}.

These institutional and individual refoundation efforts were significantly influenced by the geopolitical climate of the early Cold War era, a period marked by the political, ideological, economic, and military opposition between the Euro-Atlantic bloc and the Soviet bloc. In the early years of the Cold War, the Soviet Union pursued a policy of isolation, opting out of participation in international institutions, including those dedicated to scientific collaboration. Consequently, IUPAP predominantly consisted of members from the emerging Western bloc, supplemented by few non-aligned countries. As much as the principles accepted in 1947 declared an unconditional openness to all countries, the political landscape prevented an immediate embodiment of this ideal within IUPAP.  Moreover, while the United States was officially a national member of IUPAP, U.S. physicists were remarkably absent from the first General Assembly. As the former president of IUPAP, Robert Millikan held an ex officio membership on the Union's Executive Committee. He was therefore the liaison of the U.S. physics community with IUPAP. Ewald approached him to nominate U.S. representatives and delegates for the inaugural post-World War II General Assembly in Paris.\footnote{Ewald to Millikan, August 7, 1946, Millikan Papers, Reel 12, p. 752.} However, Millikan's advanced age precluded his attendance, and the absence of a constituted U.S. national committee of physics meant that official U.S. delegates could not be determined in time for the meeting.\footnote{Millikan to Fleury, February 28, 1947, Millikan Papers, Reel 12, p. 784; see also, IUPAP Circulaire d'Information générale, September 1947, Millikan Papers, Reel 12, p. 797.} As a result, the fifth IUPAP General Assembly (the first to convene following the war) was de facto mostly a European affair.\footnote{The only non-European physicists attending the meeting were representatives of China, and one representative from Australia. At the time both countries were Western allies, as the change of regime resulting from the victory of the Chinese Communist Party in mainland China would occur later, in 1949.}

The concurrent emergence of grassroots efforts by physicists to revitalize IUPAP and the development of a new institutional framework significantly bolstered the organization's initiatives. Prior to this period, IUPAP had not established any commissions focused on particular research areas in physics, although such provisions were included in its founding statutes. From the 1947 General Assembly onward, the formation of topical commissions emerged as a cornerstone of IUPAP's activities. This shift signified a profound transformation for IUPAP: it transitioned from an entity primarily concerned with international standardization to an organization fostering international collaboration within distinct research fields of physics. Nowhere this change is most visible that in the creation of the initial topical commissions. Notably, the earliest of these was the commission that would evolve into the Commission on Thermodynamics and Statistical Mechanics.\footnote{As discussed in the previous sections, other ICSU unions were more active in the interwar period and had already established various topical commissions. Examples are the IAU and the International Union of Pure and Applied Chemistry (IUPAC), see~\citep{blaauwHistoryIAUBirth1994,fennellHistoryIUPAC191919871994}.}

In the wake of the 1947 General Assembly, the proliferation of commissions marked a significant phase of growth for IUPAP, but the scope and function of these commissions initially lacked clarity and differed from their eventual roles. Prior to 1947, IUPAP's structure mainly consisted of general-purpose commissions, such as the SUN Commission and the Commission on Publications. The 1947 General Assembly saw the establishment of several new commissions alongside the reconstitution of the SUN Commission, which was inactive since the mid-1930s (see Table~\ref{tabCOMM}).

\begin{table}[h]
\caption{IUPAP Commissions established at the 1947 General Assembly.} 
\label{tabCOMM}%
\begin{tabular}{cll}
\toprule
Number & Name & Type \\
\midrule
I & Symbols, Unit and Nomenclature & General \\
II & International Commission for Optics & Grand commission \\
III & Thermochemistry\footnotemark[1] & Restricted \\
IV & Physico-Chemical Data  & Mixed IUPAP-IUPAC\footnotemark[2] \\
V & Viscosity & Mixed IUPAP-IUPAC-IUBS\footnotemark[3] \\
VI & Radioactive Units & Restricted \\
VII & Cosmic Rays & Restricted \\
VIII & Ionosphere & Mixed IAU\footnotemark[4], URSI\footnotemark[5], IUGG\footnotemark[6], IUPAP\\
IX & Radiometeorology & Mixed URSI, IUGG, IUPAP \\
\botrule
\end{tabular}
\footnotetext{Data taken from ``Report 5th General Assembly, September 1947," p. 5, Series B2aa, Vol. 1, IuG.
In 1947, the explicit mention of the type \textit{General} had not yet been introduced, and the types
\textit{Restricted} and \textit{Grand} were later renamed \textit{Specialized} and  \textit{Affiliated}, respectively.}
\footnotetext[1]{Soon named Thermodynamic magnitudes and notations}
\footnotetext[2]{International Union of Pure and Applied Chemistry}
\footnotetext[3]{International Union of Biological Sciences}
\footnotetext[4]{International Astronomical Union}
\footnotetext[5]{International Union of Radio Science}
\footnotetext[6]{International Union of Geodesy and Geophysics}
\end{table}

These newly established commissions fell into three categories. The first, known as ``mixed commissions," were inter-union commissions created by ICSU to tackle urgent issues by linking multiple scientific unions, including IUPAP, which ratified their creation and the participation of its members. The second category, termed ``grand commissions," focused on distinct sub-fields of physics. The first of this type was the International Commission for Optics (ICO), founded during the 1947 General Assembly.\footnote{For the foundation of the ICO, see~\citep{howardFoundingInternationalCommission2003}.} Its secretary was Pierre Fleury, a prominent French optical physicist who was also appointed secretary general of IUPAP at the same meeting. The establishment of the ICO set the precedent for the structure of grand commissions: they were independent international scientific entities that became integrated within IUPAP to pursue their objectives under its auspices. Grand commissions would immediately be renamed ``affiliated commissions," which exist to this day, but had a limited development at the time, with the second affiliated commission established only in 1974.\footnote{For the establishment of the second IUPAP affiliated commission, see~\citep{lalliBuildingGeneralRelativity2017}.} The third category, ``restricted commissions," included the Commission on Thermochemistry (Commission III), the Commission on Cosmic Rays, and the Commission on Radioactive Units. These commissions were envisioned to pursue a relatively ``limited program" compared to the broader scope of mixed and, especially, grand commissions, which aimed to forge specific physics sub-fields in the international arena. Some restricted commissions were anticipated to evolve into either mixed or grand commissions. For instance, the Commission on Radioactive Units soon transitioned into the Joint Commission on Radioactivity, a collaborative effort between IUPAP and IUPAC.\footnote{For a historical analysis of the Joint Commission on Radioactivity, see~\citep{fauqueICSUUmbrellaJoint2024}.}

What would soon evolve into the Commission on Thermodynamics and Statistical Mechanics originated from the first of the three ``restricted commissions" instituted in 1947: Commission III, initially named Commission on Thermochemistry. Its initial mission was to ``study the definition of thermochemical quantities and the notations" working in close cooperation with the SUN Commission and the corresponding commissions of IUPAC.\footnote{Fleury to Edmond Bauer, January 22, 1947, Vol. 3, Folder 20 ``Commission on Thermodynamics,'' IuQ. See also Report of the 1947 IUPAP General Assembly, Millikan Papers, Reel 12, p. 767; ``Report 5th General Assembly, September 1947," p. 7,  Series B2aa, Vol. 1, IuG.} It was soon rebranded as the Commission on Thermodynamical Magnitudes and Notations, possibly to avoid confusion with a similarly named Commission on Thermochemistry of IUPAC.  As the name indicates, the Commission's activities initially focused exclusively on standardizing symbols and units of thermodynamics at the international level. Such standardization efforts were a principal focus for the SUN Commission during the 1930s under Sir Richard Glazebrook's chairmanship.\footnote{Annex V,  ``IUPAP Procès-Verbal Quatrième Assemblèe Générale, 1934,'' p. 6, Series B2aa, Vol. 1, IuG.} Following World War II, as IUPAP aimed to resume its operations, it evidently seemed more useful to establish a specialized group of physicists solely dedicated to thermodynamics standards. This strategic decision was probably intended to harness and consolidate specific expertise in thermodynamics, rather than dispersing these efforts within the more generalized framework of the SUN Commission.

In 1947, the newborn Commission on Thermodynamic Quantities and Notations was composed of four scientists working at the intersection between physics and chemistry: French chemical physicist Edmond Bauer, serving as the provisional secretary of the Commission, alongside English physical chemist Edward A. Guggenheim, Dutch physicist and chemist Jan Hendrik de Boer, and Russian-Belgian physicist and chemist Ilya Prigogine. The inaugural meeting took place in London together with the SUN Commission and pertinent IUPAC commissions in July 1947. Although historical records detailing the deliberations of this initial gathering remain elusive, subsequent developments strongly suggest that a significant transformation of the Commission's focus was anticipated from the outset. This shift entailed a broadening of the Commission's responsibilities, culminating in a suggestion to rename it. 

In January 1948, under the auspices of IUPAP a first scientific meeting was organized by Prigogine at the Free University of Brussels, where he had been appointed professor of chemistry one year earlier. Entitled Symposium on Thermodynamics, the Brussels meeting saw the participation of 22 Western European chemists and physicists. They convened to explore topics in statistical thermodynamics, cryogenics, and irreversible processes.\footnote{``IUPAP, Minutes Sixth General Assembly (1948),''  p. 6, Series B2aa, Vol. 1, IuG.} Some of the central topics of this symposium were connected to Prigogine's own research agenda at the time, which related to the thermodynamics of irreversible processes. It also provided a first international opportunity to share knowledge resulting from wartime research that had previously not been properly disseminated.
The deliberations and outcomes of this meeting were later disseminated in the conference proceedings, published with UNESCO's support~\citep{prigogineColloqueThermodynamiqueBruxelles1949}. This event marked a foundational step in the Commission's expanded role and scope, evolving from a body primarily focused on supporting the SUN Commission's standardization efforts in thermodynamics to a thematic commission.  This new role designated it as an IUPAP commission dedicated to fostering the development of a physics sub-discipline on an international scale, primarily through the cultivation of a dedicated scientific community. The Brussels conference, as the Commission's inaugural activity, underscored this strategic pivot. The chosen approach to achieve international community-building was the orchestration of conferences centered on specialized research areas.

This modification of the Commission's functions was ratified at the sixth IUPAP General Assembly, convened in Amsterdam in July 1948, marking the second assembly after the end of World War II. At this meeting, the composition of the Commission was significantly broadened. Alongside Guggenheim, who served as president, and Prigogine, who took on the role of secretary, the original members, Bauer and de Boer, were joined by two scientists from the United States: physical chemist James Alexander Beattie from MIT and chemical physicist Joseph Edward Mayer from the University of Chicago.\footnote{``IUPAP, Minutes Sixth General Assembly (1948),''  p. 6, Series B2aa, Vol. 1, IuG.} This inclusion of U.S. scientists represented a notable shift, as they had been absent from the first post-World War II IUPAP General Assembly and so became active in IUPAP's commissions only from the Amsterdam General Assembly onward. 
 
The decisions regarding the Commission's revamped function proved to be even more pivotal. Following Prigogine's presentation of the Commission's activities, the General Assembly endorsed a significant expansion of its responsibilities and introduced a new designation: Commission on Thermodynamics and Statistical Mechanics.\footnote{Ibid., p. 13.} The narrower focus on defining magnitudes and notations within thermodynamics was agreed to continue under the collaboration with the SUN Commission. This effectively reinstated the SUN Commission's role in standardizing thermodynamic measures, as evidenced by its report at the 1948 General Assembly. Furthermore, there was an ambitious plan to host an international symposium addressing ``current problems in statistical mechanics.''\footnote{Ibid.,  p. 6.} This event was tentatively scheduled for 1949, with the location yet to be decided but under consideration for either Italy or the United States.\footnote{Ibid., p. 18.}

\section{The organization of the Florence Conference and the international (re-)construction of Italian physics after World War II}\label{sec5}

In the renewed activity of IUPAP of the immediate post-World War II period, Italy emerged as one of the most engaged national members, a status that was not assured from the outset. Given Italy's wartime alliance with Nazi Germany, there was some debate within IUPAP circles about considering Italy as an ``enemy country" in discussions concerning the participation of German physicists. For Millikan, drawing distinctions among Italy, Finland, and Germany posed a challenge, prompting a preference for evaluating participation on an individual basis rather than explicitly adopting a  policy of exclusion specifically targeting Germans alone.\footnote{Millikan to Fleury, June 18, 1947, Millikan Papers, Reel 12, p. 792.} Despite this ambiguity, the conclusion of the war repositioned Italy alongside the victorious nations. Consequently, Italian physicists were granted official participation in IUPAP starting with the 1947 Paris Assembly, contrary to Japan and Germany, whose physicists were reintegrated into IUPAP's activities only a few years later.

One could argue that one of the motivations behind Italian physicists' eagerness to collaborate with IUPAP was to improve Italy's reputation, tarnished by its role as an Axis Power during World War II.  
In the period following the war, the Italian scientific community faced major challenges such as wartime destruction and scarce resources. Additionally, the significant loss of scientists, which had begun in the 1930s, further compounded these difficulties. Many researchers had left the country, primarily for the United States, while potential young researchers often opted for different career paths. The destabilization of the scientific community began with the university professors' oath of allegiance to fascism in the 1930s~\citep{Scarantino}, and was intensified by the 1938 Fascist racial laws. At the end of the war, Italian physicists were deeply 
committed to overcoming these difficulties as they worked to rebuild
national research infrastructures and communities. Their efforts aimed
not only at restoration but also at pioneering new research practices
and initiating large-scale scientific enterprises. This attempt was
based on strong cooperation among various individual and institutional
actors and had in the process of internationalization a fundamental
component~\citep{battimelliFisiciItalianiNegli2007}.

Key figures in this (re-)construction effort were Edoardo Amaldi and Giovanni Polvani. A former student and collaborator of Enrico Fermi, Amaldi took on the task of organizing the discipline after the 1938 Racial Laws dismantled the Italian physics community, leading to the departure of many leading physicists, including Fermi, Bruno Rossi, and others. After World War II, Amaldi reaffirmed his commitment by declining a position at the University of Chicago to remain in Italy. He played a key role in rebuilding Italian physics and laid the groundwork for greater European cooperation in physics~\citep{AmaldiPhysics2000}.\footnote{Amaldi would be one of the main proponents of the European Organization of Nuclear Research (CERN) established in 1953, and of the European institutional cooperation on space research in the 1960s, see~\citep{hermannHistoryCERNLaunching1987,krigeHistoryESA2000}.}  Older than Amaldi, Polvani had been the director of the Institute of Physics at the University of Milan since 1929. During the 1930s, he transformed the institute into one of the centers for the development of modern physics in Italy. Like Amaldi, Polvani faced challenges in maintaining high-level research continuity at the institute after the 1938 diaspora and with limited resources. By focusing on cosmic-ray research, Polvani provided an opportunity to train a new generation of physicists, preparing them for the challenges of the post-World War II period~\citep{gariboldiMilanInstitutePhysics2022}.

They collaborated to reposition Italy at the heart of the systems of international cooperation that were emerging at the time. Upon his election as president of SIF in 1947, Polvani, alongside the new SIF Council, embarked on a mission to internationalize the society's leading journal, \textit{Il Nuovo Cimento}, through innovative editorial policies and ``propaganda" strategies.\footnote{``Minutes of the Meeting of the SIF Presidential Council, October 15, 1950,"  ASIF.} This drive to revitalize Italian physics through international engagement prominently featured IUPAP as a pivotal ally.\footnote{See minutes of various SIF Council's meetings in the late 1940s and 1950s, ASIF.} Italy's prompt and engaged involvement in IUPAP since the first post-World War II General Assembly led to the election of Amaldi to its Executive Committee during the second General Assembly in 1948. In turn, Amaldi's role in the IUPAP Executive Committee facilitated SIF in forging a special cooperative relationship with IUPAP. 

While the Italian committee for IUPAP was a group appointed within the CNR Committee for Physics and Mathematics, it was SIF that pushed to promote events with financial or, in some cases, even just ``moral" support from IUPAP.\footnote{``Minutes of the SIF Presidential Council Meeting, September 24, 1952," ASIF. } This support was chiefly directed towards hosting international conferences on specific research themes in Italy, an activity intimately related to the ambition of elevating \textit{Il Nuovo Cimento} to the status of an international journal and enhancing its stature within the global physics publication sphere. This is the context that led SIF to co-organize together with the Commission on Thermodynamics and Statistical Mechanics the meeting in Florence. This conference represented one of the initial and most significant efforts by SIF and Italian physicists to reassert their presence and secure a pivotal role in the realm of international scientific cooperation in the period immediately following World War II.

The suggestion to host the following conference of IUPAP Commission III (then still called Commission on Thermodynamic Notations and Quantities) in Italy was already put forward during the 1948 symposium convened by Prigogine in Brussels. Among the attendees was Roberto Piontelli, an Italian professor of electrochemistry from the University of Milan, who participated as a discussant. During the symposium, Piontelli publicly proposed that the next conference of the Commission could be held in Italy in cooperation with SIF, whose president, Polvani, was a close colleague of him at the University of Milan. Upon his return to Italy, Piontelli presented the idea to Polvani who immediately recognized its significant potential. Polvani was eager to bring the event to Italy, understanding that it would firmly integrate Italian physicists into the network of IUPAP-endorsed international scientific conferences~\citep{polvaniDiscorsoInaugurale1949a}.

Amaldi proposed hosting the conference in Florence, one of the most beautiful Italian historical cities, in May, during the ``Maggio Musicale Fiorentino," an internationally renowned opera and arts festival~\citep{polvaniDiscorsoInaugurale1949a}. After having secured the financial support of the CNR, which allocated funds to defray the local conference costs, including hotel expenses for international speakers,\footnote{Segreteria Generale del CNR to Polvani, June 30, 1948, Box 3, Folder 3, Presidenza Polvani, Correspondence 1947-1959 (hereafter Polvani-ASIF), ASIF.} SIF formally nominated Florence as the conference location. 

The enchanting Florentine setting was certainly perceived by both Italian and international physicists as an ideal backdrop to foster one of the conference's primary objectives: the formation of an international scientific community. Initial doubts as to whether to hold the conference in Italy or the United States rapidly faded away.  Florence was too attractive as a stage to establish human contacts, which, as many historians have shown, is  among the main goals of scientific conferences~\citep{biggArtGatheringHistories2023,biggCommunicatingScienceMediating2023,dastonRivalsHowScientists2023,somsenGoddessThatWe2023}. IUPAP President Henrik Kramers made it quite explicit in his inaugural discourse of the conference that Florence, as a symbol of Italian cultural renaissance, was a perfect place to support the renaissance of human interactions among physicists after the ruins of World War~\citep{kramersDiscoursOuverture1949}. As Polvani wrote to the Rector of the University of Florence, Bruno Borghi, the conference would ``be truly worthy both from a scientific point of view (the names of those who will be taking part ensure this) and from the point of view of the decorum, dignity and value that Florence and its University have unforgettably had.''\footnote{Polvani to Borghi, January 27, 1949, Carteggio e Atti dell'amministrazione centrale, Filza anno 1949, Fascicolo n.19/c, ``Convegno  Internazionale  di Meccanica Statistica,'' ASUFi, University of Florence (hereafter 19/c-ASUFi).} Other factors, such as the logistical ease for European scientists to travel to Italy compared to crossing the Atlantic and the strategic desire to reintegrate Italy into the circle of esteemed democratic and scientifically progressive nations while hastening the disassociation from its Fascist past, may have also influenced the decision. The allure of the SIF proposal was so compelling that by November 1948, the IUPAP Executive Committee not only embraced the invitation but Fleury also suggested scheduling its annual meeting in Florence immediately before the conference of the Commission on Thermodynamics and Statistical Mechanics, so that  Executive Committee members could attend the scientific conference.\footnote{Fleury to Polvani, November 27, 1948, Box 1, Folder 2, Polvani-ASIF.} 

The Italian organizers carefully prepared for the conference, aiming to showcase the venue as exceptionally beautiful and enjoyable to foster the re-establishment of professional and personal relationships following World War II. The primary venue was the National Museum of the History of Science, now known as Museo Galileo.\footnote{The conference was held in the loggia of the Castellani Palace, now the library of Museo Galileo. At that time such space was available to the homeland history delegation and it was sometimes used by the National Museum of the History of Sciences for organizing events.} 
Bruno Borghi secured the museum's grand lodge to properly ``accomodate high-profile personalities from the scientific world, including Nobel Prize winners.''\footnote{Borghi to President of the Provincial Deputation of National History, May 4, 1949, Folder ``Convegno internazionale di meccanica statistica, 1949,'' Fondo Corsini, Materiale minore Corsini I, Archive of the Institute and Museum of the History of Science, now Museo Galileo (hereafter AIMSS).}\footnote{While Borghi used the plurals, the Florence conference would be attended by one Nobel Laureate at the time (Wolfgang Pauli). Three of the speakers would receive the Nobel Prize after the conference, one in physics (Max Born) and two in chemistry (Lars Onsager and Ilya Prigogine).}
This effort was part of a broader initiative to elevate the venue's prestige and highlight Italy's significant contributions to modern science's evolution. Among the notable displays was one of the most celebrated portraits of Galileo by Justus Sustermans, on loan from the Galleria Palatina for the conference's duration.\footnote{Soprintendenza alle Gallerie per le Province di Firenze, Arezzo e Pistoia, Verbale della provvisoria consegna del ritratto di Galileo Galilei al Museo delle Scienze di Firenze, May 17, 1949, AIMSS.} The museum's director and staff dedicated themselves fully to the event's success.\footnote{Polvani to Corsini, May 27, 1949, AIMSS.} Conference attendees enjoyed complimentary access to Florence's major galleries and museums, such as the Galleria degli Uffizi and Galleria Pitti. They were also treated to a classical music concert, an artistic tour, and, unsurprisingly, high-quality local cuisine, both at the venue and in the charming suburb of Fiesole.\footnote{Polvani to Borghi, April 15, 1949, 19/c-ASUFi; Borghi to Azienda Autonoma del Turismo, June 5, 1949, AIMSS; and Invitation to the Gala lunch at the Aurora restaurant in Fiesole, May 17, 1949, 19/c-ASUFi.} The dinner, in particular, was highlighted by Polvani for its role in offering Italian participants tangible benefits and a unique opportunity to share a communal space with international guests.\footnote{Polvani to Borghi, April 15, 1949, 19/c-ASUFi.}

\begin{figure}[h]%
\centering
\includegraphics[width=0.5\textwidth]{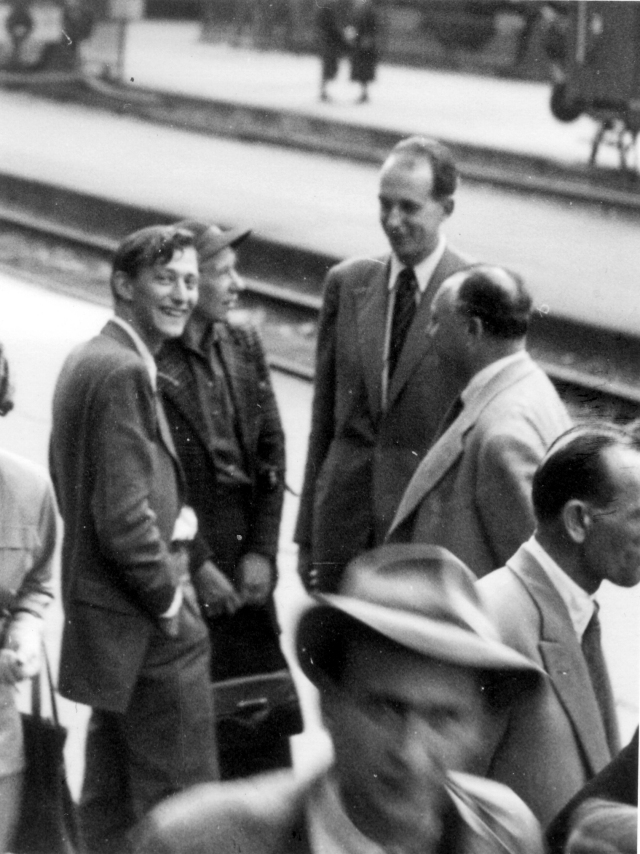}
\caption{Return journey from the Florence Conference, probably at the central railway station of Milan. 
From left to right: Joaquin M. Luttinger, Franca Pauli, Max R. Schafroth and Wolfgang Pauli.
Courtesy CERN, Geneva, PAULI-ARCHIVE-PHO-073.}
\label{Pauli}
\end{figure}

The strategic role of internationalization in the reconstruction of Italy's national physics community is  illustrated through three key initiatives undertaken by SIF in conjunction with the Florence conference. First, this international gathering served as a catalyst for SIF to secure funding from the University of Florence and other local entities, aimed at providing two substantial grants. These grants were designed to support extended research stays, lasting no less than 90 days, for two early-career SIF members at prestigious institutions abroad. Dubbed ``Borse Città di Firenze," these scholarships were established by the Rector of the University of Florence. The allocated funds for each grant were notable for the era, amounting to 500,000 Italian Lire apiece (approximately equivalent to 10,000 Euros in today's currency) \citep{BandoDiConcorso1949}.\footnote{For the revaluation of currencies, we used the calculation tool in https://rivaluta.istat.it/.} SIF's strategy of leveraging conferences as opportunities to secure local funding for research grants had previously proven effective during national congresses. However, the international prominence of the Florence conference enabled SIF to significantly enhance the success of this initiative, securing a remarkable increase in funds. This signals the pivotal role of the Florence conference in advancing Italian physics, particularly in facilitating international training. The conference's international stature and the distinction of its attendees, including the 1945 Nobel Laureate Wolfgang Pauli (Fig.~\ref{Pauli}), were instrumental in highlighting the event's significance and in drawing substantial support for the field within Italy. 

Initially, Polvani conceived the Florence research grants as a strategic initiative to bolster theoretical physics studies broadly. He proposed not to limit the research topics sponsored by the grant to statistical mechanics, envisioning that this choice would have open more opportunities to young Italian physicists.\footnote{Polvani to Borghi, January 27, 1949, 19/c-ASUFi.} While at least one SIF board member supported this view,\footnote{Rostagni to Polvani, April 30,1949, Box 3, Folder, Polvani-ASIF.} further discussion within the SIF Council led to reconsider this initial goal. In its final wording, the call declared that the grants were designed to enhance the recipients' expertise in theoretical physics, with a particular emphasis on statistical mechanics. Eventually, the funds were allocated especially to support research trips tied to statistical mechanics, showing the commitment to advance Italian scholarship in this specialized area. 
One of the two recipients was Giorgio Careri, who was supported his studies on the presence of a statistical order in the condensed phase, conducted in collaboration with Mayer at the Institute for Nuclear Studies at the University of Chicago.\footnote{Giorgio Careri to Presidenza della Società Italiana di Fisica, July 26, 1949, Box 2, Folder 2; Careri to Polvani, July 15, 1950; and  Joseph E. Mayer to Italian Physical Society, Box 3, Folder 2, Polvani-ASIF. Careri's research stay in Chicago was further supported by a grant of the Fulbright Program obtained with the support of Amaldi~\citep[pp. 173-84]{bonolisMaestriAllieviNella2008}. Careri's research with Mayer was published in~\citep{mayerEquationStateComputations1952}.} A few years after his return, Careri established schools focused on the properties of liquid helium and low temperature physics, topics that were novel in the Italian physics landscape~\citep[pp. 113-43]{bonolisFisiciItalianiTempo2003}~\citep[see also][]{bonizzoniNascitaFisicaDlela2002}. This initiative significantly enhanced Italian research in areas related to statistical mechanics and condensed matter physics. In later recollections, Careri himself emphasized the pivotal role of the Florence conference in the opening up of these new research avenues in Italy~\citep{careriLarsOracle2000,careriStudiSperimentaliDi2003}.\footnote{The second recipient was Giampiero Puppi; see ``Minutes of the SIF Presidential Council Meeting, September 3, 1949," ASIF. However, it in unclear whether Puppi actually made use of the funds and, eventually, his research interests remained strictly related to theoretical particle physics.}

The second line of action was closely linked to the strategy of internationalization of the society's journals. From the outset, it was planned that the papers from the conference would be featured in a UNESCO-sponsored special issue, to be published in \textit{Il Nuovo Cimento}'s supplement, known as \textit{Supplemento al Nuovo Cimento}.\footnote{This was the issue number 2 of \textit{Supplemento al Nuovo Cimento} of Vol. 6 of \textit{Il Nuovo Cimento} published in 1949, see https://link.springer.com/journal/40761/volumes-and-issues/6-2/supplement.} The SIF Council saw the special issue as a way to strengthen and quicken the growth and international standing of \textit{Il Nuovo Cimento}.\footnote{See, ``Minutes of the SIF General Assembly, November 9, 1948,'' ASIF.} Polvani, along with other SIF Council members, championed this internalization effort despite facing opposition from some of the society's more traditionalist members, who lamented the excessive use of English in the flagship journal of the Italian Physical Society.\footnote{``Minutes of the SIF General Assembly, September 16, 1950;'' and ``Minutes of the SIF Council Meeting, October 15, 1950,'' ASIF.}

The third and perhaps more subtle initiative involved the concurrent planning of a separate event: an international conference on cosmic rays, scheduled for September 1949 in Como, Alessandro Volta's birthplace, to celebrate the 150\textsuperscript{th} anniversary of the invention of the Voltaic pile. This conference was sponsored by the second IUPAP topical commission established in 1947, the Commission on Cosmic Rays. From its inception, the conference was envisioned to complement an IUPAP-sponsored international congress in Basel,  so that the joint venture could attract to Como many ``world-famous non-Italian scientists."\footnote{``Minutes of the SIF General Assembly, November 9, 1948,'' ASIF.} Especially notable was the anticipated attendance of Enrico Fermi, who had never returned to Europe after leaving Italy in 1938~\citep{LaNazione}.
Given Fermi's planned visit to Italy and Europe, the SIF Council regarded the Como conference as potentially surpassing the Florence conference in significance and scope.\footnote{During his 1949 trip, Fermi gave a number of lectures on atomic physics in Milan and Rome. For the English translation of of these lectures with commentaries, see~\citep{tucciEnricoFermiLectures2024}.} By the end of 1948, SIF began collaborating with IUPAP on both the Florence and Como conferences, enhancing their mutual impact and showcasing the Italian physics community as exceptionally active in hosting IUPAP-affiliated international scientific gatherings. The strategic linkage and synergy between these two conferences, as revealed in Polvani's uncatalogued correspondence stored in the SIF archives, indicate their integral role in the broader strategy to internationalize Italian physics, with IUPAP serving as a key partner during this period.\footnote{See, numerous documents dated 1948/49 in Polvani-ASIF.}

These activities underscore the complex interplay of strategies deployed by leading Italian physicists to elevate the stature of Italian physics through concerted internationalization efforts. The Florence conference emerged as a pivotal moment in this ambitious campaign, marking the first significant step by SIF and the Italian physics community to reassert and reclaim a prominent position within the international scientific arena in the aftermath of World War II. This strategy proved so effective that by 1950, Amaldi could proudly declare that ``Italy was the favored nation" for hosting IUPAP-endorsed international conferences and congresses in the immediate postwar era, a testament to the success of Italian physicists' approach.\footnote{``Minutes of the SIF Council Meeting, September 16, 1950,'' ASIF.}

\section{The role of the Florence conference in an emerging research field}\label{sec6}

\begin{figure}[h]%
\centering
\includegraphics[width=0.49\textwidth]{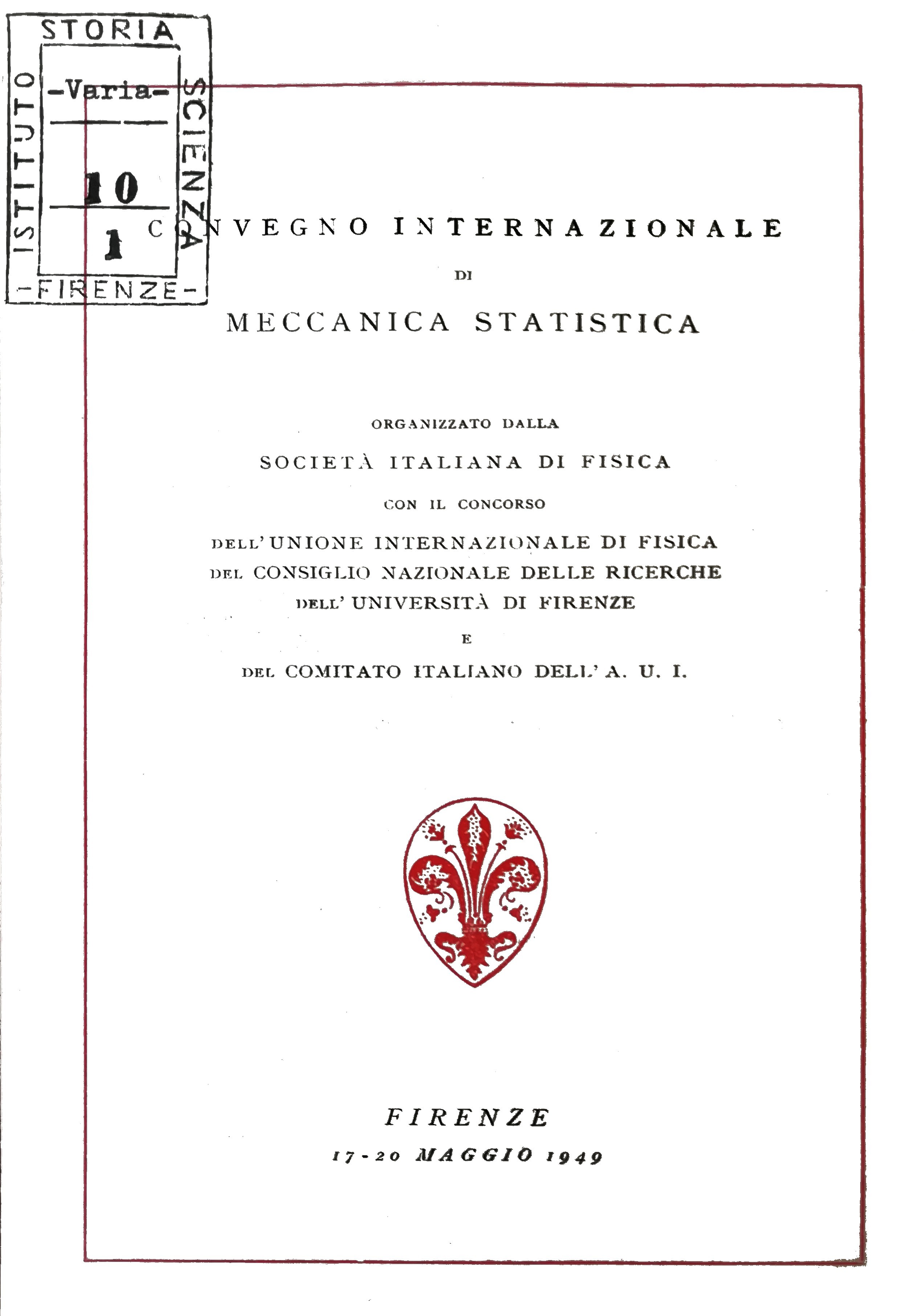}
\includegraphics[width=0.49\textwidth]{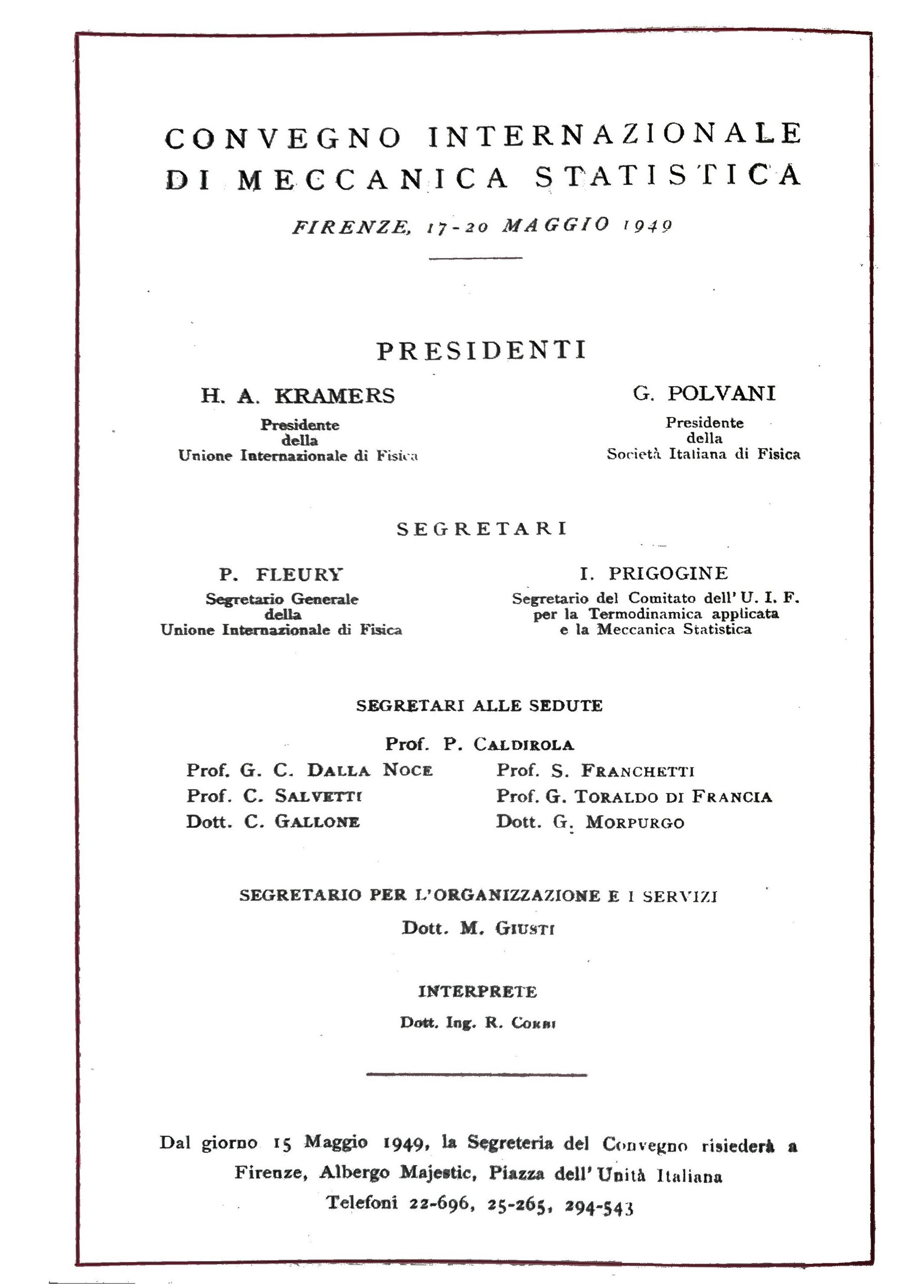}

\includegraphics[width=0.49\textwidth]{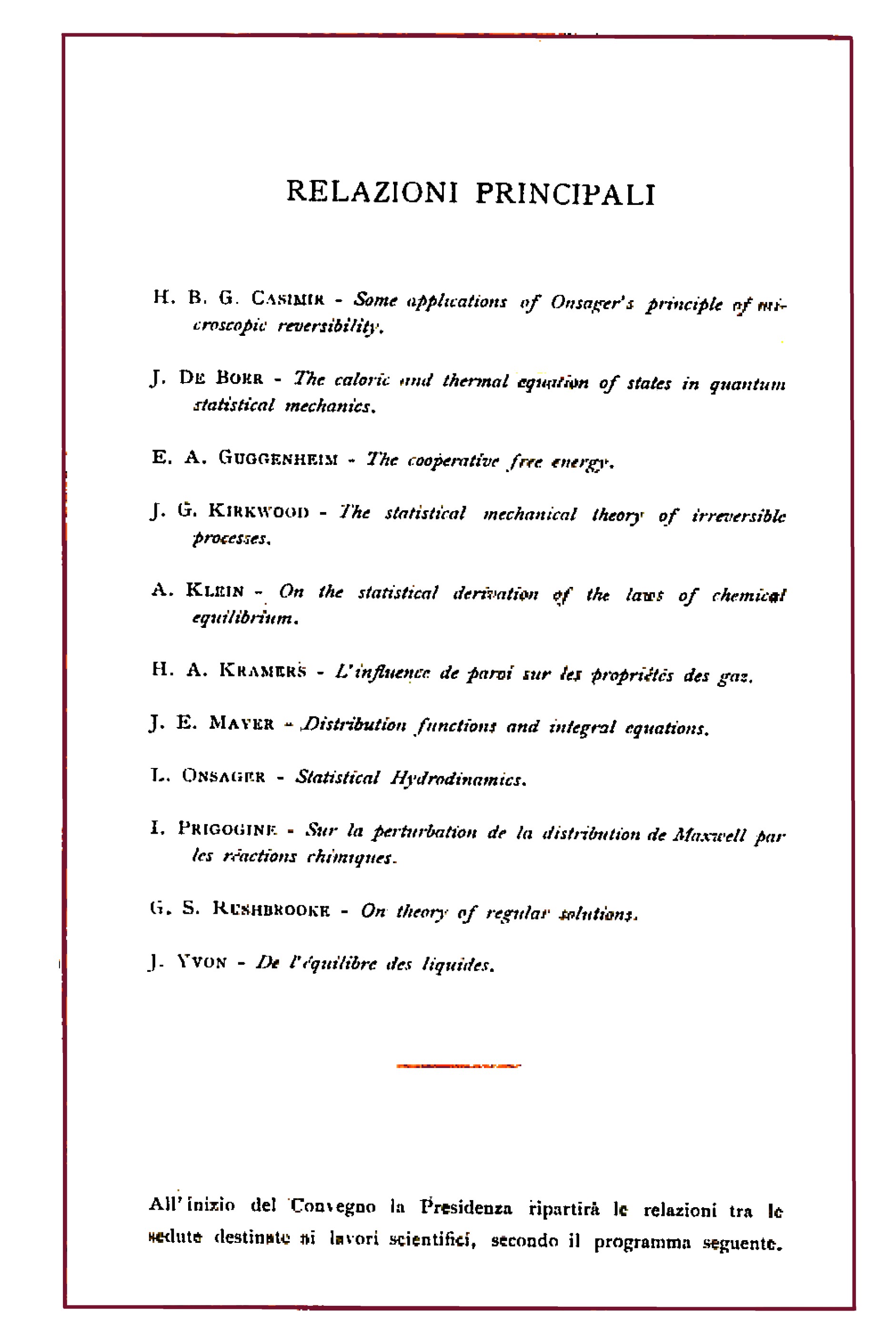}
\includegraphics[width=0.49\textwidth]{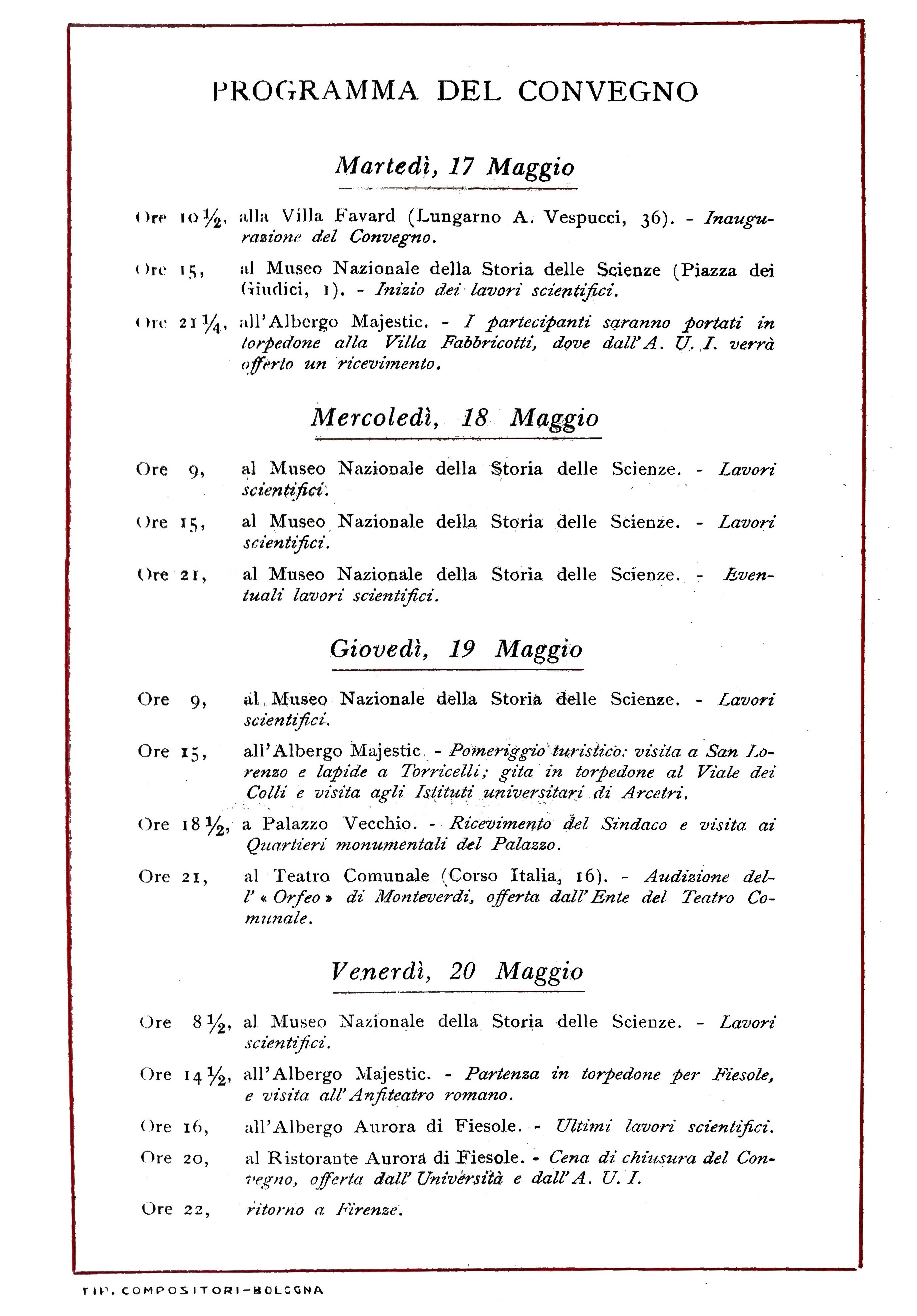}
\caption{Conference program. Folder ``Convegno internazionale di meccanica statistica, 1949,'' Fondo Corsini, Materiale minore Corsini I, AIMSS.}
\label{programma}
\end{figure}

\begin{figure}[h]%
\centering
\includegraphics[page=2,width=0.49\textwidth,trim=1.5cm 4cm 1.5cm 1cm,clip=true]{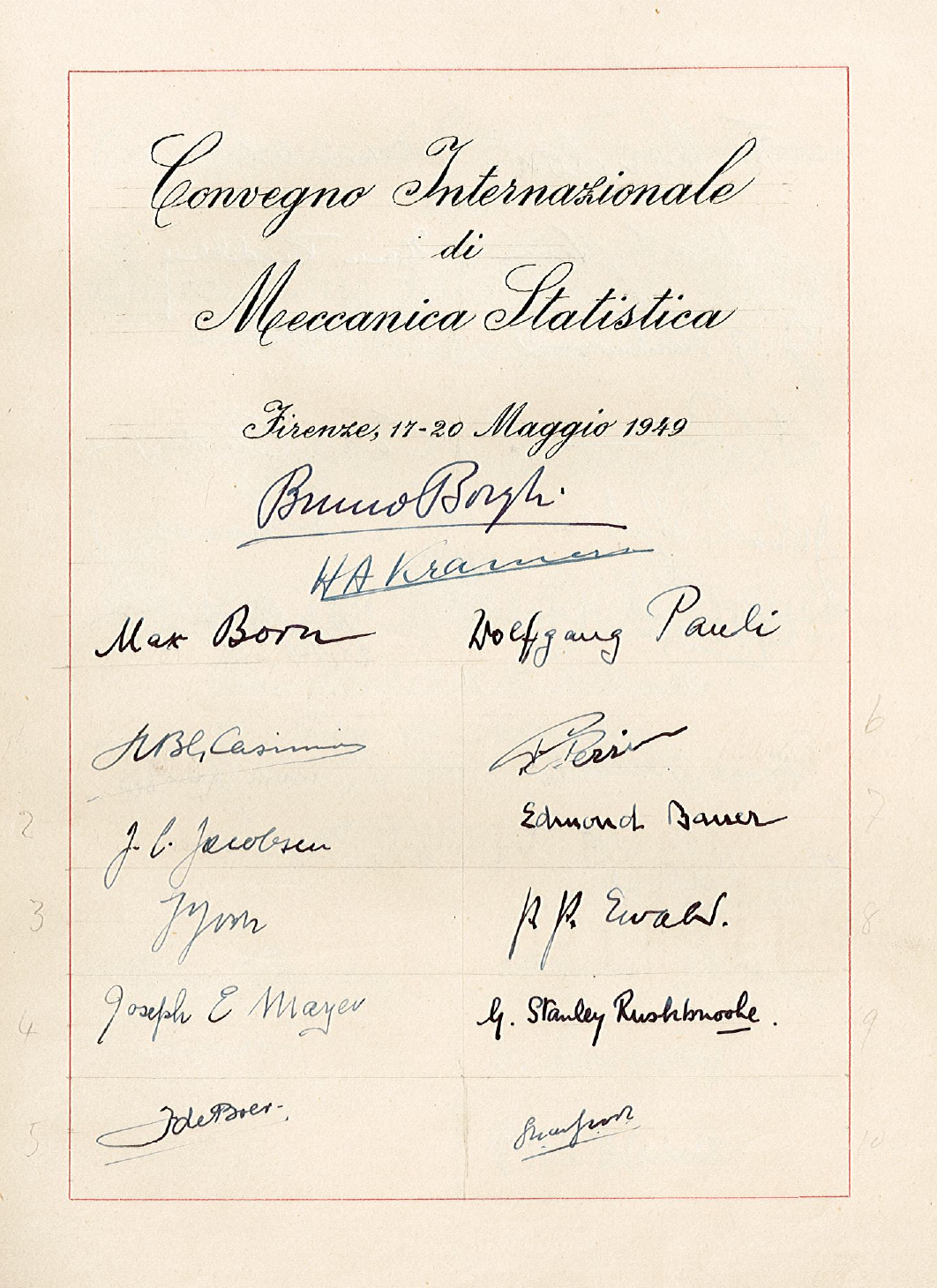}
\includegraphics[page=3,width=0.49\textwidth,trim=1cm 1.5cm 2cm 3.5cm,clip=true]{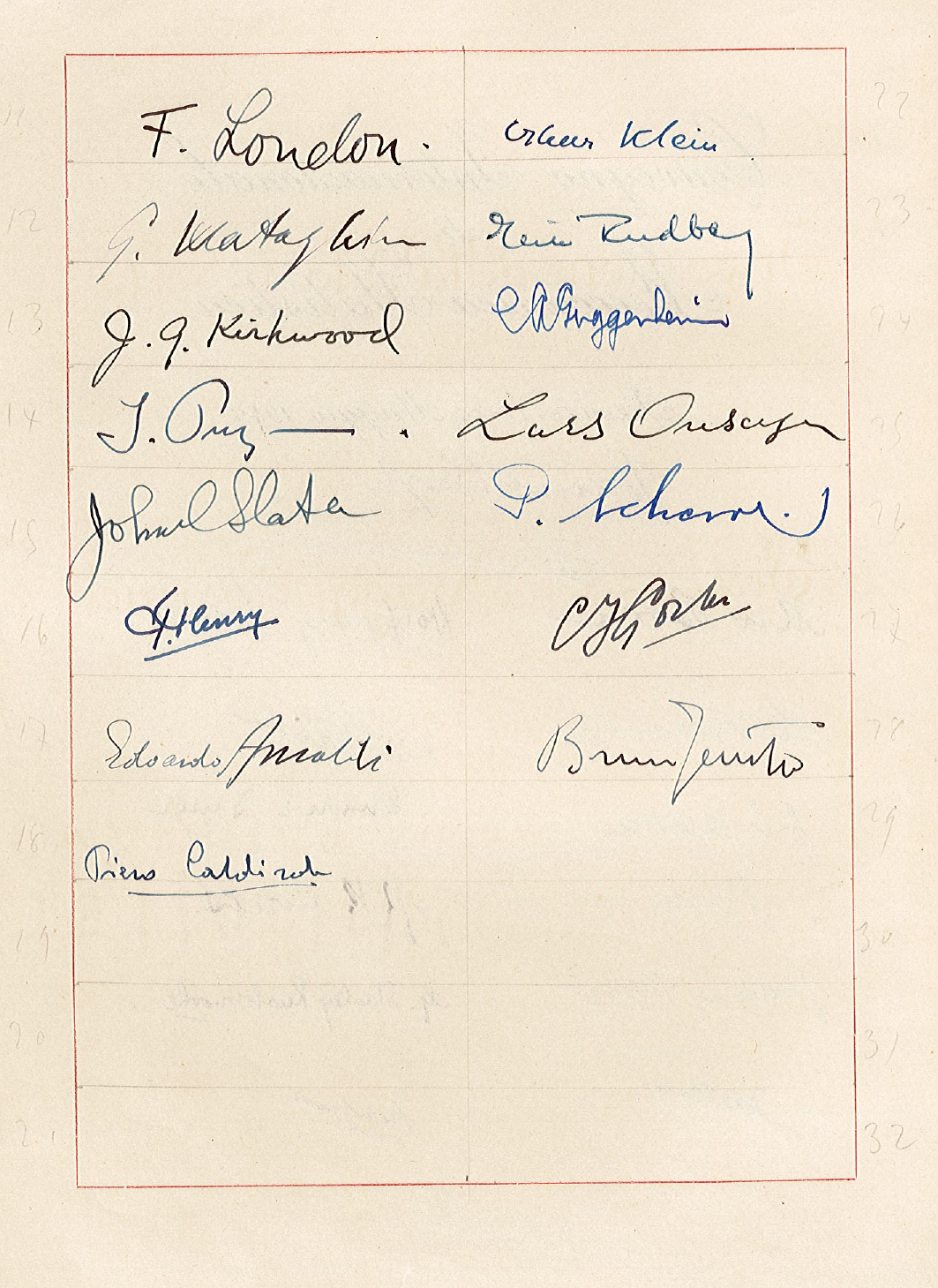}
\caption{Attendance sheet of the International Congress on Statistical Mechanics, 17-20 May, 1949, ASIF 
(names are reported as they are written).
Left page, from top to bottom and from left to right:
Bruno Borghi, H.A. Kramers, Max Born, Wolfgang Pauli, H.B.G. Casimir, F. Perrin,
J.C. Jacobsen, Edmond Bauer, J. Yvon, P.P. Ewald, Joseph E. Mayer, G. Stanley Rushbrooke,
J. de Boer, Garikian.
Right page, from top to bottom and from left to right:
F. London, Oskar Klein, G. Wataghin, Erik Rudberg, J.G. Kirkwood, E.A. Guggenheim,
I. Prigogine, Lars Onsager, C. Slater, P. Scherrer, P. Fleury, C.J. Gorter, Edoardo Amaldi, Bruno Zumino,
Piero Caldirola. 
}
\label{foglio_firme}
\end{figure}

The conference took place over four days, from May 17 to 20, 1949, with the opening ceremony hosted at Villa Favard, at that time the auditorium for the Faculty of Economic and Commercial Sciences of the University of Florence, and the scientific sessions predominantly conducted at the National Museum of the History of Science (see program in Fig.~\ref{programma}).\footnote{Borghi, Press release, International Conference on Statistical Mechanics, May 16, 1949, 19/c-ASUFi.} The last scientific session and the closing gala dinner occurred at the Albergo Ristorante Aurora in Fiesole. Official records indicate a participation of seventy to eighty attendees, with a notable attendance sheet featuring signatures of distinguished participants (Fig.~\ref{foglio_firme}).

While SIF took care of most logistical aspects of the conference, the scientific agenda reflected the efforts of the IUPAP Commission on Thermodynamics and Statistical Mechanics to establish its objectives and scope following its 1948 rebranding. 
This event, being the Commission's first scientific gathering post-renaming during the IUPAP General Assembly in Amsterdam, aimed to outline its thematic focus and affirm its commitment to primarily organizing and endorsing conferences.

The program's formation was a collaborative endeavor involving Prigogine, the Commission's secretary; Hendrik Kramers, the president of IUPAP and co-chairman of the conference; and Polvani, the president of SIF as well as the conference's co-chairman. Since its inception, the gathering was named international conference on ``statistical mechanics" departing from the Commission's initial focus on thermodynamics as illustrated by the 1948 Brussels symposium. 
Under the umbrella name of ``Statistical Mechanics,''\footnote{According to Martin Klein~\citep{Klein1990},
this term was coined by J. Willard Gibbs in his 1884 presentation at a Philadelphia meeting of 
the American Association for the Advancement of Science~\citep{Gibbs1884}.} 
the organizers had in mind a focus on the statistical mechanics of interacting systems articulated in five thematic sub-categories that were shared with invited speakers: mathematical methods for determining the configuration integral;\footnote{Also known as 
partition function.} 
cooperative phenomena; distribution functions; the statistical mechanics of irreversible phenomena; and the quantum foundations of statistical mechanics.\footnote{Polvani to Mr. Arnoux, March 24, 1949, Box 3, Folder 2, Polvani-ASIF, see also Italian text of invitation attached to Polvani to Borghi, February 8, 1949, 19/c-ASUFi.}  This thematic structuring aimed at stimulating in-depth discussions across vital frontier areas within statistical mechanics, showcasing the evolving ambitions of the Commission.

In order to contextualize the conference's role in sub-discipline formation from a scientific standpoint, 
let us briefly discuss the status of statistical mechanics at the time, noting that 
historical reconstructions of its development throughout 
the 20\textsuperscript{th} century are still incomplete.\footnote{To our knowledge, the most comprehensive accounts of the long-term development of statistical mechanics are the contributions by Cyril Domb~(\citeyear{dombThermodynamicsStatisticalMechanics1995a})
and Max Dresden~(\citeyear{dresdenNonEquilibriumStatisticalMechanics1995}) in~\citep{brownTwentiethCenturyPhysics1995c}} At the start of the century several scientific sub-fields were interconnected not only because some scientists contributed to multiple domains, but also because it was possible to borrow tools and concepts from other fields.\footnote{From this point of view, statistical mechanics has always been strongly interdisciplinary.} A notable example is the statistical considerations used to study black body radiation, which led Max Planck to formulate his quantum hypothesis~\citep{badinoBumpyRoadMax2015}.

The 1930s marked a transitional period in both education and research. Statistical ensembles were known from the era's onset, but Fowler's 1936 book on statistical mechanics~\citep{fowlerStatisticalMechanicsTheory1936}
 mentions the canonical ensemble only in passing, while a few years later,
Schr\"odinger significantly contributed to the dissemination of Gibbs' ideas~\citep{dombThermodynamicsStatisticalMechanics1995a}.
This period also saw significant advances in the study of phase transitions, perhaps the most important and
succesful domain in statistical physics.
The existence of transitions between different equilibrium phases of a substance with varying 
temperature or pressure was well-known, as was the law of
corresponding states, an early example of \textit{universality}~\citep{landauStatisticalPhysics1938}.
However, precise experiments and Lars Onsager's 1944 exact solution of the two dimensional Ising model~\citep{onsager1944crystal} were needed 
to convince the scientific community that the characterization of a phase transition as described by the van der Waals equation was incorrect.

By the time of the Florence conference, the understanding of equilibrium phenomena was advanced enough to guide research and clarify several open problems~\citep{van1957statistical}: finding techniques to determine the partition function, developing a satisfactory theory
of liquids, and moving beyond the classical theory of phase transitions, to name a few. In contrast, nonequilibrium phenomena  were much less understood~\citep{dresdenNonEquilibriumStatisticalMechanics1995}. In addition, they covered such a wide range
of phenomena that it was (and still is) impossible to encompass them within a single theory.
After the formulation of the Boltzmann equation, a significant advancement was made by Onsager, who described transport processes in systems kept not too far from equilibrium,
leading to the development of ``nonequilibrium thermodynamics"~\citep{onsagerReciprocalRelationsIrreversible1931}.

Systems far from equilibrium can exhibit a completely different phenomenology. In this domain, Prigogine and his Belgian school were paying a pivotal role. They presented their approach at the 1948 Brussels symposium and in the 1949 Florence.\footnote{Prigogine's work in this period laid the groundwork for his 1960s research on the role of dissipative structures in thermodynamic systems far from equilibrium, which would earn him the Nobel Prize in Chemistry in 1977; for a review of Prigogine's theory, see~\citep{prigogineTheoryDissipativeStructures1973}.} However, Prigogine encountered considerable skepticism when he presented his work on irreversible phenomena at the Brussels symposium. He recalled that his presentation was met with hostility, for the prevailing view among experts was that statistical mechanics should focus on equilibrium states, viewing irreversible processes as merely transitory phenomena~\citep{prigogineBiography1993}.\footnote{If we trust Prigogine's recollection, we might assume that, by designating irreversible phenomena as one of the five central topics of the Florence conference, Prigogine, with the support of Kramers, aimed to challenge this prevailing view and position the study of irreversible phenomena as as cutting-edge research agenda within the community being fostered through such gatherings.}

More generally, after World War II, the field of statistical mechanics was not only fragmented into various research trajectories but also deeply intertwined with the evolution of emerging research fields, whose development varied significantly across different countries. In particular, it was  connected to the growing research sub-fields of chemical physics and solid-state physics. Regarding Italy, more specifically, statistical mechanics was still in its infancy, with little to no established research agendas actively advancing the field~\citep{careriStudiSperimentaliDi2003}.

To capture how the Commission was reconfiguring its focus and scope through the Florence conference it is instructive to make a comparison with the 1948 symposium held in Brussels. The Florence conference notably surpassed the Brussels symposium in international participation, attracting nearly triple the number of attendees and featuring speakers from eight diverse countries (see Appendix~\ref{app_lista}), as opposed to the Brussels symposium, where speakers hailed predominantly from five Western European nations, with the majority originating from just Belgium (7) and The Netherlands (4) \citep[see list of participants in][]{prigogineColloqueThermodynamiqueBruxelles1949}. 
Despite the number of speakers being very similar in the two conferences (around fifteen),
the Florence event demonstrated a significantly broader geographical distribution. This expansion in reach, while primarily within the Euro-Atlantic sphere akin to the IUPAP membership of that era, nonetheless marked a considerable enlargement of the Commission's international engagement. Notably, the Florence conference included participants from two continents and featured at least one delegate from the emerging Eastern bloc, the Polish physicist Jan Weyssenhoff, underlaying the conference's role in broadening the Commission's international footprint.

Scientifically, the distinct emphases of the two meetings are clearly shown by their titles. The Brussels symposium was dedicated to thermodynamics, in contrast to the Florence conference, which centered on statistical mechanics. This distinction goes much beyond differences between titles. It was significantly reflected in the themes of the presentations, as documented in their respective proceedings.
The presentations at Brussels were signaling that the main focus of research concerning the Commission was statistical thermodynamics. The animated discussions following most presentations at the Brussels symposium showed that many of the employed assumptions and approaches were matter of contention. This was particularly evident in the opening scientific talk,``The statistical basis of thermodynamics,"  which explored the foundation of thermodynamic laws on a unique statistical principle \citep{guggenheimStatisticalBasisThermodynamics1949}.\footnote{Guggenheim was also present at the Florence conference, where he gave a talk discussing the Boltzmann factor \citep{guggenheimCooperativeFreeEnergy1949}.} Given by the president of the Commission, Edward Guggenheim, the opening talk of the symposium suggested a collective lean towards recognizing statistical mechanics as a vital area for developing an international research community. While in the Brussels' proceedings the lion's share is taken up by the thermodynamics of open systems promoted by Prigogine and the locals, in Florence several contributions  focused on more fundamental problems of statistical mechanics~\citep{bornFoundationQuantumStatistics1949,kleinStatisticalDerivationLaws1949,guggenheimCooperativeFreeEnergy1949,deboerQuantumPropertiesCondensed1949,kirkwoodStatisticalMechanicalTheory1949};
this passage accompanied a more basic Commission's shift in the scientific direction, prefigured by its renaming. As discussed in the Section~\ref{sec4}, moving from Commission on Thermodynamic Magnitudes and Notations to Commission on Thermodynamics and Statistical Mechanics had already marked the progression toward assuming a scientific role rather than a purely technical one as previously envisaged. This strategic shift to statistical mechanics, seen as closer to the heart of physics and offering a richer vein of research possibilities, was evident in the content featured at the Florence conference in contrast to the Brussels symposium.

The Commission's shift toward statistical mechanics was operationalized by the Florence conference through the delineation of the five above-mentioned sub-topics. This strategic focus marked the Commission's evolution beyond its traditional thermodynamics realm, aiming instead to anchor itself within the physics community by spearheading the exploration of specific, cutting-edge research areas. The commitment to these sub-topics signified an intention not just to serve as a community hub but to actively shape the research landscape. This ambition was realized as the proceedings of the conference showcased: all five sub-topics were comprehensively addressed by the speakers, albeit with a variable number of papers dedicated to each sub-topic.%
\footnote{For the mathematical methods of the configuration integral, see \citep{montrollContinuumModelsCooperative1949};
cooperative phenomena were addressed in \citep{guggenheimCooperativeFreeEnergy1949,rushbrookeTheoryRegularSolutions1949};
distribution functions were the topic of \citep{mayerDistributionFunctionsIntegral1949,prigoginePerturbationDistributionMaxwell1949}; 
the statistical mechanics of irreversible phenomena was explicitly tackled by 
\citep{casimirAspectsOnsagerTheory1949,kirkwoodStatisticalMechanicalTheory1949,wataghinIrreversibleProcessesFormation1949}; 
and the quantum foundations of statistical mechanics was the topic of Born's lecture \citep{bornFoundationQuantumStatistics1949}.} 
Such thorough coverage reaffirms the conference's success in redirecting the attention of the community around the Commission towards statistical mechanics. 
The opening speech of Kramers~(\citeyear{kramersDiscoursOuverture1949}) openly reveals the intention to position statistical mechanics as a field with historical roots in the kinetic theory of gases, now poised to incorporate the recent advances in quantum physics and to confront fundamental and outstanding problems, including the derivation of the thermodynamic properties of a system from the interaction among its elementary constituents, starting from the comprehension of ferromagnetism and the condensation of a gas.

Some presentations at the conference have been recognized, a posteriori, as influential on the later developments of what was then considered an emerging field. This is particularly manifest in Lars Onsager's lecture, titled \textit{Statistical Hydrodynamics}. His paper published in \textit{Supplemento al Nuovo Cimento}~\citep{onsagerStatisticalHydrodynamics1949} has amassed over 2,000 citations to date, dwarfing the citation counts of other papers from the proceedings 
by two orders of magnitude.\footnote{This analysis was made using Google Scholar. A research on Scopus confirms the later impact of the paper with a count of 1362 citations from 1970 to 2024 with an evident increase of citations starting from the late 1990s.} Onsager's paper studies the statistics of a system of vortices, that represents the first and one of the most important examples displaying (absolute) negative temperatures. In it, Onsager, without providing a full demonstration, introduced the hypothesis of quantized vortical lines in superfluid helium as a hydrodynamic invariant. Remarkably, at the time, the significance of this work was not fully recognized, just as it is notable that the discovery for which Onsager was awarded the Nobel Prize  for Chemistry in 1968~\citep[his reciprocal relations identified in][]{onsagerReciprocalRelationsIrreversible1931,onsagerReciprocalRelationsIrreversible1931a}
was the focus of Casimir's presentation at the Florence Conference~\citep{casimirAspectsOnsagerTheory1949}, rather than his own. Only after this conjecture was reformulated by Feynman in the 1950s and, consequently, put to test, the powerful anticipation by Onsager at the Florence conference was fully recognized~\citep{feynmanChapterIIApplication1955}.%
\footnote{It is also worth noting a short comment of Onsager (him again) to the paper by Rushbrooke (\citeyear{rushbrookeTheoryRegularSolutions1949}, p. 261),
where Onsager gives the analytical expression of the magnetization of the two dimensional Ising model, a result which will be published (together with its independent derivation)  three years later by C.N. Yang~(\citeyear{yang1952spontaneous}).}

Evaluating the immediate impact of the Florence conference on the field of statistical mechanics proves challenging.  Here, we limit ourselves to a simple bibliometric approach, examining the occurrence of bi-grams such as 
\textit{statistical physics, statistical mechanics,} and \textit{phase transition(s)} 
in \textit{The Physical Review}, considered the most important physics research periodical 
of the time and representative of mainstream research trends ~\citep[see][]{khelfaouiPhysicalReviewPeriphery2019}: 
in the decade following the conference (1950-1959), the frequency of these terms is significantly higher than 
in the decade immediately before (1939-1948).\footnote{By frequency 
we mean the fraction of articles where the string appears. More precisely, the frequency in the full articles increases
by approximately 50$\%$ and the occurrence in titles alone is more than doubled.} 
While this does not necessarily imply a direct causative relationship between the conference and the field's growth, it at least indicates that the Commission successfully aligned its research focus towards a field burgeoning at the time. The Florence conference, coupled with the renaming of the Commission, marked it as a catalyst for community building, mirroring developments within related physics sub-disciplines during that era: in 1947, a similar identification process within the physics community led to the establishment of the solid-state division of the American Physical Society, largely promoted by industrial physicists~\citep{martinSolidStateInsurrection2018,weartSolidCommunity1992}. Within IUPAP, the engagement of industrial physicists at that time was comparatively minimal, steering the emphasis towards statistical mechanics as a discipline ripe for fostering and supporting an international scholarly community with the Florence conference clearly being the genetic event for this community.

\section{The solidification of a tradition from topical conferences to StatPhys}\label{sec7}

The Florence conference served as a defining moment, heralding a new era for the Commission on Thermodynamics and Statistical Mechanics, while also establishing precedents for other topical commissions. From that moment on, the Commission embraced its principal role as the sponsor of international conferences. This transition to a formalized communication routine was swiftly institutionalized. The initial 1949 gathering paved the way for a succession of regular conferences, starting with the conference on phase transitions held in Paris in 1952, after which these gatherings became a staple in the scientific community's calendar, occurring biennially or triennially as outlined in Table~\ref{tabSP}. Remarkably, less than a decade following the Florence conference, Italy once again played host to one of these regular international conferences organized by the Commission. This was the conference on \textit{Condensed states of simple systems}, which convened in Varenna in 1957.

The shifting venues of these conferences illustrate the geopolitical dynamics inherent to the development of IUPAP as a whole. Over time, there has been a marked expansion in the geographical diversity of host countries, extending well beyond the Western European character of the early post-World War II era. The admission of the Soviet Union into IUPAP in 1957 was particularly significant in this respect, as it facilitated East-West cooperation amidst the Cold War. Furthermore, the period saw an enhanced engagement of scientific communities from non-Western countries, reflecting a shift towards inclusivity in the global scientific arena during the post-colonial era.\footnote{For studies of the geopolitical changes in IUPAP after 1957, see \citep{cozzoliEdoardoAmaldiScientific2024,hofParticlesPurityPolitics2024,lalliDiplomacyPhysicsBack2024,liuGettingOutNegative2024,olsakovaIUPAPCooperativeAntagonism2024,dasilvanetoSocialistInternationalismScience2024,simonRestoringPhysicsIUPAP2024,turchettiOnlyTenseEncounter2024}.}

It is beyond the scope of this article to discuss the scientific and political processes characterizing the almost three decades of activities of the Commission on Thermodynamics and Statistical Mechanics from the Florence conference up to the 1977 conference in
Haifa, when the designation `StatPhys' was officially adopted. Instead, this section aims to illuminate key transitions within the Commission's activities and its conferences. These transformations not only cemented and elucidated the Commission's role but also revealed how its members conceptualized its identity and functions during a crucial period in its history.

\begin{table}[h]
\caption{StatPhys conferences}
\label{tabSP}%
\begin{tabular}{@{}cll|cll@{}}
\toprule
Edition & Where  & When & Edition & Where & When \\
\midrule
1   & Florence (Italy)\footnotemark[1]  & 1949  & 16 & Boston (USA) & 1986   \\
2  & Paris (France)\footnotemark[2]  & 1952  & 17  & Rio de Janeiro (Brazil)  & 1989  \\
3  & Brussels (Belgium)\footnotemark[3]  & 1956  & 18  & Berlin (Germany)  & 1992  \\
4  & Varenna (Italy)\footnotemark[4]  & 1957 & 19  & Xiamen (China) & 1995  \\
5  & Utrecht (Netherlands)\footnotemark[5] & 1960  & 20  & Paris (France) & 1998  \\
6  & New York (USA)\footnotemark[6]  & 1962  & 21  & Cancun (Mexico) & 2001  \\
7  & Aachen (Federal Republic of Germany)\footnotemark[7]  & 1964  & 22  & Bangalore (India) & 2004  \\
8  & Copenhagen (Denmark) & 1966  & 23  & Genoa (Italy)  & 2007  \\
9  & Kyoto (Japan)  & 1968  & 24  & Cairns (Australia) & 2010  \\
10  & Chicago (USA)  & 1971  & 25 & Seoul (South Korea) & 2013  \\
11 & Amsterdam (Netherlands)  & 1973  & 26  & Lyon (France)  & 2016  \\
12   & Budapest (Hungary)  & 1975 & 27 & Buenos Aires (Argentina) & 2019  \\
13   & Haifa (Israel)\footnotemark[8]  & 1977  &28  & Tokyo (Japan)\footnotemark[9] & 2023  \\
14    & Edmonton (Canada) & 1980 & 29 & Florence (Italy)\footnotemark[10] & 2025\\
15 & Edinburgh (UK) & 1983 & & & \\
\botrule
\end{tabular}
\footnotetext{This table was compiled using data sourced from the IUPAP reports, Series B2aa ``General Reports,'' Vols.1-3, IuG. The denomination StatPhys was officially introduced in 1977 to designate the Haifa conference as StatPhys13. By assigning this sequence number, the Commission implicitly applied the StatPhys label to all prior meetings in the series, thereby retrospectively recognizing them as part of the StatPhys conference lineage, as shown in the table.}
\footnotetext[1]{\it International conference on statistical mechanics}
\footnotetext[2]{\it Phase transitions}
\footnotetext[3]{\it Transport phenomena}
\footnotetext[4]{\it Condensed states of simple systems}
\footnotetext[5]{\it Many body problems}
\footnotetext[6]{\it Theory of phase transitions}
\footnotetext[7]{{\it Statistical mechanics of equilibrium and non-equilibrium}. From this edition the conference
is no more focused on a specific issue.}
\footnotetext[8]{From this edition onwards the conferences take on the official name StatPhys and take place every three years.}
\footnotetext[9]{Postponed due to COVID-19 pandemic.}
\footnotetext[10]{Scheduled.}
\end{table}

Between 1952 and 1962, the conferences focused on distinct research topics, as indicated in Table~\ref{tabSP}. It was not until 1964 that the conferences began to resemble their current format, adopting a broader focus on statistical mechanics as a whole, rather than delving into more narrowly defined areas of inquiry within this broader domain. This shift likely stemmed from an aspiration to transform the conference into a regular gathering point for the entire global community of scientists working on statistical physics. The goal was to foster a more inclusive forum, rather than one catering exclusively to specialized topics of interest to particular sub-communities, regardless of their size.

The 1970s witnessed significant transformations concerning the policies and structures of IUPAP commissions and their sponsored conferences. Starting from the 1969 General Assembly in Dubrovnik, IUPAP embarked on a comprehensive revision of policies related to the composition of commissions and the criteria for conference sponsorship. In response to the increasing volume of conference proposals from IUPAP commissions, the Executive Committee of IUPAP introduced a set of evaluation criteria focused on scientific merit, international representation, and organizational solidity. A member of the U.S. national committee, William Havens Jr., further contributed to these deliberations by proposing a taxonomy for classifying conferences into three distinct categories, which received tentative approval during the 13\textsuperscript{th} General Assembly in Dubrovnik.
The type-A conferences, called ``general conferences," were designed ``to provide an overview of the entire field of interest to a Commission, and would normally occur at three-year intervals"  with a foreseeable attendance of about 750-1500 scientists. Type-B conferences were called ``topical conferences" and were designed to ``concentrate on broad sub-fields" within the general area of interest of the Commission. The foreseen attendance was about 300-600 physicists. The smaller Type C conference were called ``special conferences" and ``would concentrate on much more restricted specialized topics" than the type-B conferences. The expected attendance was in the range of 50-200 physicists.\footnote{\textit{IUPAP, Report 1970}, esp. p. 25, p. 29, pp. 81-83, Series B2aa, Vol. 2, IuG.} At the subsequent IUPAP General Assembly, held in Washington D.C. in 1972, new procedural guidelines were established for the composition of commissions and the election of their members. Incidentally, a new identification system for commissions was adopted, denoted as ``C.
followed by a sequential number reflecting their chronological establishment, a system that remains in use today.''\footnote{\textit{IUPAP, Report 1973}, pp. 14-18, Series B2aa, Vol. 2, IuG.}\footnote{The full point after C was later eliminated and currently IUPAP commissions are simply denoted by C plus a sequential number, with C3 being the denomination of the Commission on Statistical Physics.}

The reorganization within IUPAP had significant implications for the operations of all its commissions. In 1972, the designation of the Commission on Thermodynamics and Statistical Mechanics as C.3 cemented its reputation as the inaugural topical commission of IUPAP, following C.1, the Commission on Finances, and C.2, the general-purpose SUN Commission. Prompted by these general reconfigurations within IUPAP, in 1973 the Commission resolved to inaugurate an award recognizing  outstanding achievements in thermodynamics and statistical mechanics. The newly appointed chairman, Canadian physicist Donald D. Betts, announced to the Commission members that the concept of the award was met with enthusiastic endorsement, both within the Commission and among the wider physics community. Betts advocated for the award to take the form of a gold medal rather than a cash prize, arguing that it would not only be economically feasible for the Commission but also offer a lasting symbol of achievement. Following a Commission's vote on names of founding fathers of the sub-discipline, the award was eventually called the Boltzmann Medal. The inaugural Boltzmann Medal was presented at the 1975 conference on Statistical Mechanics to Kenneth Wilson, recognizing his pioneering work on the renormalization group. This decision came despite Betts' earlier communication in 1973, which had suggested Lars Onsager as the ``obvious choice"\footnote{Betts to Members of the Commission on Thermodynamics and Statistical Mechanics, March 9, 1973, see also Betts to Larkin Kerwin, September 19, 1973, Series E3  ``Correspondence of the Commissions,'' Vol. 2, Folder C 3, IuG.} for his contributions to the study of irreversible processes, for which he had already been awarded the Nobel Prize in Chemistry in 1968.\footnote{Lars Onsager died in 1976. It was therefore not possible to take him into consideration as a recipient of the next Boltzmann Medals. Later, in 1982, Kenneth Wilson also won the Nobel Prize in Physics. The same occurred to Giorgio Parisi who received the Boltzmann medal in 1992 and the Nobel prize in 2021.}

The second significant shift in the mid-1970s aimed to establish a more systematic scheduling of the Commission on Thermodynamics and Statistical Mechanics' international conferences, proposing a biennial interval for these conferences starting in 1975.\footnote{Betts to Kerwin, September 19, 1973, Series E3  ``Correspondence of the Commissions,'' Vol. 2, Folder C 3, IuG.} This initiative was in line with the above-mentioned broader initiative to categorize IUPAP-sponsored international conferences, as endorsed during the 1969 IUPAP General Assembly. The conferences organized by Commission C.3 naturally fell into the Type-A category, embodying the essence of a \textit{general} conference. The decision in the early 1960s to forgo specific titles in favor of a more inclusive approach was exactly in this direction. However, aligning the Commission's ambition for biennial conferences with IUPAP's organizational framework proved challenging. The financial and administrative constraints of adhering to a two-year cycle were compounded by IUPAP's triennial budgetary cycle, which spanned from one General Assembly to the next. As Larkin Kerwin, the IUPAP secretary general, explained to Betts, supporting Type-A conferences biennially within the confines of a three-year budget was problematic.\footnote{Kerwin to Betts, October 22, 1973,  Series E3, Vol. 2, Folder C 3, IuG.} Additionally, during this period, IUPAP officials urged all commissions to reassess their roles in light of recent disciplinary advancements. Commission C.3 was tasked to provide a report in 1977 ``examin[ing] the boundaries of its field and to consider whether these should be modified, whether the Commission should be divided, fused with another commission, or dissolved" as well as to reassess its field through the lens of its international conferences.\footnote{Kerwin to Herbert B. Callen, October 10, 1975, Series E3, Vol. 2, Folder C 3, IuG.} This directive underscored the need for commissions to remain adaptive and responsive to the evolving landscape of their respective disciplines.

This request initiated a self-reflection among the members of Commission  C.3 regarding its activities, roles, and functions. In 1977, the chair of the Commission in 1975-78, U.S. statistical physicist Herbert B. Callen, submitted a preliminary report to Kerwin, articulating the Commission members' perception of both the Commission and the field it served during that time. It is beneficial to quote extensive passages from this report, as they directly illustrate the views of Commission's members about the Commission itself and its scientific domain within the broader landscape of physics.

``I believe that I reflect the consensus among statistical physicists
when I assert a unique status for statistical physics among the varied
subdisciplines of physics. That uniqueness derives from the \textit{generality} of
statistical physics; it is at the same time pervasive throughout physical
theory, and absent [as] a private, wholly-owned domain.\footnote{These statements can be embraced even after almost fifty years.
On the one hand, this attests to a positive peculiarity of statistical physics.
On the other hand, this can lead to problems of recognition within the broader community of physicists.} Organizationally we find
our Commission to be the second oldest in IUPAP (and the first Commission
dealing with a substantive subdiscipline), and we also find that statistical
physics is the least intensively organized of all subdisciplines (witness
the absence of a Statistical Physics Division within the American Physical
Society). Ours is therefore an integrative role, unifying an international
community of statistical physicists many, or even most, of whom count
themselves simultaneously as members of one or another subdiscipline of
physics. This interdisciplinary linkage has marked the field itself as well as
its organizational structure, as some of the most significant contributions of
statistical mechanics have transcended traditional boundaries. Perhaps the most
recent and dramatic examples of this have been in the fields of broken symmetry
and in the development of the techniques of the renormalization group."

This view of the Commission's role and function was accompanied by a definition on the field itself:
``[w]hereas the sub-atomic disciplines of physics probe for deeper structures
below the atom scale, the macroscopic disciplines explore the collective
properties of large aggregates of interacting atoms. The approach can be
particularistic, as in the theory of the solid state or plasma physics,
or it can be general. Thermodynamics and statistical mechanics constitute the study of those
properties of macroscopic systems which follow from general model-independent
laws of physics. In practice, additionally, statistical mechanics often
encompasses the exploratory application of those principles to particular
models of macroscopic matter."\footnote{Callen to Kerwin, July 26, 1977, Series E3, Vol. 2, Folder C 3, IuG. Emphasis in the original.}

The comprehensive report of the Commission, which also chronicled its history and the strategies deployed to fulfill its objectives,
especially through the organization of the regular
international conferences on statistical mechanics,
was presented to the IUPAP Executive Committee shortly afterwards. The Executive Committee wholeheartedly endorsed the Commission's endeavors and reaffirmed its continued operation. Nonetheless, this review process resulted in certain adjustments that further solidified the Commission's operational framework. Starting in 1977, the conferences were scheduled to occur triennially, aligning with IUPAP's three-year budgetary cycle. Additionally, a new official name for the conference series was adopted. Henceforth, the general A-type conferences on statistical mechanics organized by the Commission were to be known as StatPhys.\footnote{\textit{IUPAP, General Report 1979}, pp. 54-55, Series B2aa, Vol. 2, IuG; and P. C. Hemmer, International conferences on statistical mechanics, 1973-1977, November 10, 1978, folder 4,21 ``Commission Thermodynamics and Statistical Mechanics,'' Larkin Kerwin fonds (P202), subseries P202/B4 IUPAP, Division de la gestion des documents administratifs et des archives, Universit\'e Laval, Quebec, Canada.}

\section{Conclusion}\label{sec8}

By designating the 1977 Haifa conference as StatPhys13, the Commission's members implicitly established a lineage of StatPhys conferences, acknowledging the 1949 Florence conference as StatPhys1. This recognition positioned the Florence conference as the inaugural event in a long-standing tradition of international community-building activities within a specific physics sub-discipline. This decision marked the official acknowledgment of the Florence conference's fundamental importance at a critical juncture in the Commission's history, a period characterized by a reevaluation of its role and scientific scope. As discussed in Section~\ref{sec4}, the Florence conference was not the Commission's first organized scientific event; it was preceded by the 1948 Symposium on Thermodynamics in Brussels, which focused on statistical thermodynamics. The choice to recognize the Florence conference as the series' commencement, overlooking the Brussels symposium, sheds light on the Florence event's significance in the collective memory of the community associated with the IUPAP conferences on statistical mechanics. This may suggest that the Brussels conference had been forgotten altogether by the Commission's members or that, due to its more limited geographical reach and narrower thematic scope, it was deemed not representative of the StatPhys conference tradition's origins. Regardless of the thematic variances between the Brussels and Florence meetings, the 1977 decision emphatically highlighted the Florence conference's role as the genesis of a scientific tradition embodied by the StatPhys conferences.

However, while acknowledging the historical actors' perspectives on their sub-disciplinary tradition, we have demonstrated the significant, albeit distinct, role played by the Brussels conference. This earlier gathering enabled members of the Commission at that time to engage in discussions that led to a reevaluation of the Commission's role, culminating in a decision to rename it and broaden its scope. Considering its pivotal role in setting the stage for the subsequent Florence conference, we suggest that the Brussels conference be recognized within the StatPhys series as StatPhys0. This approach aligns with practices adopted for labeling other international conferences, e.g., those on General Relativity and Gravitation.\footnote{When the GR label was introduced, the 1957 Chapel Hill conference was retroactively designated GR1, while the 1955 Bern conference was identified as GR0, see \citep{lalliBuildingGeneralRelativity2017}.} Such a distinction would not diminish the Florence conference's significant contributions to community building and field definition. Instead, it would offer a more comprehensive and nuanced account of Commission C3's history.

In our paper, we elucidated the multiple functions of the Florence conference, one of which gained formal recognition nearly three decades later when it was retrospectively designated as StatPhys1. The conference marked the beginning of the Commission on Thermodynamics and Statistical Mechanics' efforts to forge an international community of physicists dedicated to emerging themes within a sub-discipline, a research field that, as Callen highlighted in 1977, lacked institutional representation at the national level. More comprehensively, the Florence conference was instrumental in redefining the Commission's primary role to include the organization of international conferences aimed at both fostering a global community and advancing the sub-discipline. This initiative set a precedent together with the meetings of the Commission on Cosmic Rays (Commission C.4 from 1972) and thus became a model for the goals and activities of all subsequent IUPAP topical commissions. 

The community-building impact of the Florence conference was further magnified by the fact that, for many attendees, it was their first opportunity to convene since the devastation of World War II. The endeavor to move beyond the war's atrocities and engage in a new communal experience with international peers was likely intensified by the event's setting in Florence, a city celebrated as a jewel of the Renaissance. The organizers intentionally chose this location for its symbolic representation of the rejuvenation of international collaboration post-war, thereby amplifying the conference's significance in the collective memory and fostering a spirit of renewal among the scientific community.\footnote{All these aspects emerge clearly from the opening speeches given by
Borghi~(\citeyear{borghiParoleDiSaluto1949}), Kramers~(\citeyear{kramersDiscoursOuverture1949}), and Polvani~(\citeyear{polvaniDiscorsoInaugurale1949a}).}

The conference held significant importance at the local level as well, serving as the inaugural step in a strategic endeavor by Italian physicists to revitalize Italian physics post-World War II through internationalization and to open up new research fields beyond cosmic rays as well as nuclear and subnuclear physics, the traditional workhorses of the Italian physics community. The Florence conference was the first scientific event co-sponsored by IUPAP and the Italian Physical Society, marking the beginning of SIF's forceful campaign to establish itself as a key collaborator with IUPAP in hosting international scientific conferences. This series of initiatives not only facilitated the promotion of Italian physics on the international stage but also played a crucial role in internationalizing its premier publication, \textit{Il Nuovo Cimento}. Consequently, \textit{Il Nuovo Cimento} ascended to prominence as a leading journal for particle physics in Europe in the subsequent years, reflecting the successful intertwining of local revitalization efforts with the broader objective of international engagement~\citep[see][]{lalli_crafting_2021}.

Revisiting the taxonomy of international scientific conferences recently introduced in~\citep{biggArtGatheringHistories2023}, the Florence conference aligns with the category of disciplinary conferences, assuming that this category is broad enough to encompass gatherings focused on sub-disciplinary fields. This adaptation reflects the trend observed in the latter half of the 20\textsuperscript{th} century, when the expansion of disciplinary communities led to an increase in conferences dedicated to specialized sub-disciplines. While disciplines had replaced nations as the primary denominator of scientific conferences in the early 20\textsuperscript{th} century, as argued in~\citep{biggArtGatheringHistories2023}, the post-World War II era saw sub-disciplines gradually taking precedence as the focal point of such events, a shift vividly illustrated by the Florence conference. This conference not only served as a benchmark for an emerging sub-discipline in identifying and deliberating on cutting-edge research agendas but also provided a platform for defining the scientific and epistemological underpinnings essential for cultivating a distinct community of physicists, thereby differentiating it from broader scientific communities focused on thermodynamics. In this sense, the viewpoint that the primary purpose of scientific conferences is to celebrate the community through the performance of rituals, as suggested by~\citep{somsenGoddessThatWe2023}, is only partially corroborated by this case study. A critical function of the conference was to lay the groundwork for a sub-discipline, setting the stage for the development of a specialized community.  Additionally, an element of science diplomacy was implicitly at play, as the collaboration between IUPAP and SIF symbolized Italy's full integration into the international scientific cooperation network under IUPAP's Euro-Atlantic framework, an integration that occurred prior to other international conferences being held in Italy and before initial discussions regarding the establishment of a European laboratory for high-energy physics research commenced.

The success of the Florence conference significantly bolstered the Commission on Thermodynamics and Statistical Mechanics, setting a precedent for future IUPAP commissions. This pivotal event initiated a period of growth and self-identification for the Commission, as detailed in Section~\ref{sec7}. The transformative years of the 1970s solidified the Commission's identity, a foundation that has remained stable to the present day. Subsequent developments within the Commission can be viewed as natural progressions from these well-established roots. The focus on statistical mechanics, initiated at the Florence conference and further emphasized by the subsequent rebranding of the Commission's general conferences as the StatPhys series, culminated in a significant renaming of the Commission itself in 1990 to the ``C.3 Commission on Statistical Physics.''\footnote{C.3 Commission on Thermodynamics and Statistical Mechanics, in Reports of International Commissions of IUPAP and the Inter-Union Committees presented to the 20\textsuperscript{th} General Assembly, Dresden- September 1990, pp. 5-7, on p. 7; see also, \textit{IUPAP, General Report 1990}, Appendix B, p. 15 , Series B2aa, Vol. 3, IuG.} This modification was due to the understanding that, while statistical physics was initially related to the thermodynamics of fluids and magnets, from the 1970s the spectrum of research topics addressed within statistical physics became much broader, including ``random media, dynamical systems, soft condensed matter, chaos, growth phenomena, and fractals.'' As noted by the 1993 Commission C.3 report, ``statistical physics has moved into more complex problems, which are closer to applications in diverse fields like biology, neural networks, computer science, astronomy or geophysics.''\footnote{ C.3 Commission on Statistical Physics, in Reports of International Commissions of IUPAP and the Inter-Union Committees presented to the 21\textsuperscript{th} General Assembly, Nara, Japan - September 1993, pp. 9-11, on p. 9, Series B2aa, Vol. 3, IuG.} The most recent major change was introduced less than twenty years ago when IUPAP started financing a Young Scientist Prize for each commission with the same three-year periodicity of general conferences.\footnote{For the list of awardees of Commission C3 Young Scientist Awards, see, https://archive2.iupap.org/commissions/c3-commission-on-statistical-physics/c3-awards/.}

The oversight of StatPhys conference organization and the bestowment of Boltzmann and Young Scientist awards remain the primary duties of the C3 Commission to date, a role that has only been accentuated by the Commission's resolution to almost exclusively endorse type-A conferences.\footnote{In the 2010s only a few type-B conferences were scheduled and StatPhys is the only type-A 
conference organized by C3. Private communication by Stefano Ruffo, former C3 chair, to one of the authors (PP) and
information obtained from the C3 Commission's website, https://iupap.org/who-we-are/internal-organization/commissions/c3-statistical-physics/.} This strategic decision underscores the significance of StatPhys as the only conference championed by the C3 Commission for the whole community. The selection process for conference venues adheres to a rotational principle across continents, with the choice of country and city determined from submitted applications. There is no invitation procedure, with the decision being dictated by the desire to valorize an existing and growing local activities in statistical physics. In 2016, StatPhys26 in Lyon was the occasion to create a weekly newsletter in Statistical Physics, which has since informed the entire community about events and job offers. While this newsletter might seem a minor activity, it is still a direct offshoot of the StatPhys conferences, further demonstrating the long-lasting impact of a tradition inaugurated by the Florence conference.

\backmatter


\bmhead{Acknowledgments}

The text is partially based on research done by historians of science participating
in the project ``One Hundred Years of IUPAP: A History," coordinated by Jaume
Navarro and one the authors (RL). We are very grateful for the support
of IUPAP, which funded the digitization of its historical archive and the
project's workshops, which this paper has also benefited from. 
We are pleased to acknowledge the invaluable support of our colleagues
Giovanni Battimelli, Adele La Rana, Giovanni Paoloni and Stefano Ruffo for their comments and suggestions. 
We are also very grateful to the following individuals who helped us in our research of archival and historical materials: Antonella Cotugno (Archives of the Department
of Physics at Sapienza University), 
Francesca Gallori (Biblioteca Riccardiana, Florence),
Leonardo Gariboldi (Archivio Polvani, University of Milan),
Antonella Gasperini (Archivio Storico Osservatorio Arcetri),
Erika Ghilardi (Archivio storico Foto Locchi),
Alessia Glielmi (CNR Archives, Rome)
Nicolien Karskens (Leiden University Libraries), 
Alessandra Lenzi (Museo Galileo, Florence),
Anna Ranzi (Casa Museo Villa Monastero),
Fioranna Salvadori (Archivio di Deposito e Storico, University of Florence),
Maria Valdez (Documentary Information Services, Royal Netherlands Academy of Arts and Sciences),
and Veronica Vestri (Deputazione di storia patria per la Toscana).
We want to express our deep gratitude to two anonymous referees for their deep reading of our paper and the insightful comments they provided. 
Finally, we are extremely grateful to the president of the Italian Physical Society,
Angela Bracco, its honorary president, Luisa Cifarelli, and the society's staff, especially Barbara Alcarani e Barbara Alzani, for the
permission to access and cite the historical materials stored in the SIF
offices in Bologna and their support during the research.

\section*{Declarations}

\begin{itemize}
\item Funding

PP acknowledges support from the MIUR PRIN 2017 project 201798CZLJ.

\item Conflict of interest/Competing interests (check journal-specific guidelines for which heading to use)

We declare no competing interests. 
\item Ethics approval 

Not applicable.
\item Consent to participate

Not applicable.
\item Availability of data and materials

Not applicable.
\item Code availability 

Not applicable.
\item Authors' contributions

All authors contributed equally to the study conception and design. Data collection and analysis were performed by both authors  in different archives. The first draft of the Sections~\ref{sec1}, \ref{sec2}, \ref{sec3}, \ref{sec4}, \ref{sec5} was written by Roberto Lalli, the first draft of the Sections~\ref{sec6}, \ref{sec7}, \ref{sec8} was written by Paolo Politi. All authors commented on previous versions of the manuscript. All authors read and approved the final manuscript.

\end{itemize}


\begin{appendices}

\section{List of participants}%

\label{app_lista}
This is the list of participants as reported by Polvani to Borghi on May 7, 1949. In a couple of cases (E. Bauer and C. Slater), expected keynote speakers didn't give a lecture at the conference
(or at least no account appears in the Proceedings).
\footnote{See, Polvani to Borghi, May 7, 1949, 19/c-ASUFi.} The presidents of the conference were H.A. Kramers and G. Polvani.

\subsection{Members of the Executive Committee of the International Physical Union}

We also indicate the position held within IUPAP.\newline
H. A. Kramers (President, Leiden)\nl
E. Amaldi (Rome)\nl
P. Auger (Unesco representative, Paris)\nl
P. P. Ewald (Belfast)\nl
P. Fleury (Secretary, Paris)\nl
C. J. Gorter (Leiden)\nl
J. C. Jacobsen (Copenhagen)\nl
P. Scherrer (Zurich)\nl
C. Slater (Cambridge, Massachusetts).

\subsection{Keynote speakers}

We also indicate the titles of the scientific contributions, as reported in the proceedings.\newline
E. Bauer (Paris)\newline
M. Born (Edinburgh) \textit{The foundation of quantum statistics}\newline
H. B. G. Casimir (Eindhoven) \textit{Some aspects of Onsager's theory of reciprocal relations
in irreversible processes}\nl
J. de Boer (Amsterdam) \textit{The caloric and thermal equation 
of states in classical and in quantum statistical mechanics}\nl
E. A. Guggenheim (Reading) \textit{Co-operative free energy}\nl
J. C. Kirchwood (Pasadena) \textit{The statistical mechanical theory of irreversible processes}\nl
O. Klein (Stockholm) \textit{On the Statistical Derivation of the Laws of Chemical Equilibrium}\nl
H. A. Kramers (Leiden) \textit{On the behaviour of a gas near a wall}\nl
J. E. Mayer (Chicago) \textit{Distribution functions and integral equation method}\nl
E. W. Montroll (Washington) \textit{Continuum models of cooperative phenomenon}\nl
L. Onsager (New Haven) \textit{Statistical Hydrodynamics}\nl
W. Pauli (Zurich) \textit{Conferenza (fuori programma) sulla Elettrodinamica quantistica}\nl
I. Prigogine (Secretary, Brussels) \textit{Sur la perturbation de la distribution de Maxwell
par des reactions chimiques}\nl
G. S. Rushbrooke (Oxford) \textit{On the theory of regular solutions}\nl
C. Slater (Cambridge, Massachusetts)\nl
I. Yvon (Strasbourg) \textit{De l'equilibre des liquides}.

\subsection{Invited}

We also indicate the position held within the conference (if any), as reported in the official program.
In two cases (C.J. Gorter and G. Wataghin) we also report the titles of their scientific contributions.
\vfill
\begin{minipage}{0.45\textwidth}
G. Alvial (Santiago)\nl
E. Amaldi (Rome)\nl
D. Baroncini (Bologna)\nl
G. Boato (Rome)\nl
G. Bolla (Milan)\nl
G. B. Bonino (Bologna)\nl
A. Borsellino (Milan)\nl
P. Caldirola (Chairman, Pavia)\nl
G. Careri (Rome)\nl
N. Carrara (Florence\nl
R. Casale (Roma)\nl
E. Clementel (Padua)\nl
E.D.G. Cohen (Amsterdam)\nl
G.C. Dalla Noce (Chairman, Bologna)\nl
V. De Sabbata (Bologna)\nl
L. Fabbrichesi (Padua)\nl
B. De Finetti (Trieste)\nl
S.R. De Groot (Utrecht)\nl
S. Franchetti (Chairman, Florence)\nl
S. Gallone (Chairman, Milan)\nl
Garikian (Brussels)\nl
Gehenieau (Brussels)\nl
M. Giusti (Secretary, Florence)\nl
L. van Hove (Brussels)\nl
C.J. Gorter (Leiden) \textit{The two fluid model for Helium II}\nl
R. Jost (Z\"urich)\nl
R. Jastrow (Leiden)
\end{minipage}
\hfill
\begin{minipage}{0.45\textwidth}
E. Keberle (Bern)\nl
J. van Kranendonk (Amsterdam)\nl
A. Loinger (Pavia)\nl
F. London (Durhen, North Carolina)\nl
P. Marcus (London)\nl
G. Morpurgo (Chairman, Rome)\nl
F. Perrin (Paris)\nl
P. Pinto (Milan)\nl
L. Prinzi (Padua)\nl
G. Puppi (Padua)\nl
L. Rolla (Genoa)\nl
L. Rosino (Bologna)\nl
A. Rostagni (Padua)\nl
E. Rudberg (Stockholm)\nl
C. Salvetti (Chairman, Milan)\nl
R. Schafroth (Z\"urich)\nl
G. Semerano (Padua)\nl
P. Straneo (Genoa)\nl
G. Todesco (Parma)\nl
G. Toraldo di Francia (Chairman, Florence)\nl
P. Udeschini (Milan)\nl
G. Valle (Bologna)\nl
M. Verde (Z\"urich)\nl
G. Wataghin (Turin) \textit{Irreversible processes and the formation of nuclei}\nl
J. Weyssenhoff (Cracovia)\nl
G. Zin (Turin)\nl
B. Zumino (Rome)
\end{minipage}




\end{appendices}
\end{document}